# Operando Methods and Probes for Battery Electrodes and Materials

Alex Grant[1] and Colm O'Dwyer[1,2]*

[1]School of Chemistry, and Environmental Research Institute, and Tyndall National Institute, University College Cork, Cork, T12 YN60, Ireland

[2]Advanced Materials and BioEngineering Research Centre, Trinity College Dublin, Dublin 2, Ireland

## Abstract

With the importance of Li-ion and emerging alternative batteries to our electric future, predicting new sustainable materials, electrolytes and complete cells that safely provide high performance, long life, energy dense capability is critically important. Understanding interface, microstructure of materials, the nature of electrolytes and factors that affect or limit long term performance are key to new battery chemistries, cell form factors and alternative materials. The electrochemical processes which cause these changes are also difficult to probe because of their metastability and lifetimes, which can be of nanosecond to sub nanosecond time domains. Consequently, developing and adapting high-resolution, non-destructive methods to capture these processes proves challenging, requiring state-of-the-art techniques. Recent progress is very promising, where optical spectroscopies, synchrotron radiation techniques, and energy-specific atom probe tomography and microscopy methods are just some of the approaches that are unravelling the true internal behaviour of battery cells in real-time. In this review, we overview many of the most promising non-destructive methods developed in recent years to assess battery material properties, interfaces, processes, and reactions under operando conditions in electrodes and full cells.

email: c.odwyer@ucc.ie

**1. Introduction**

Substantial progress in battery technology is essential if we are to succeed in an energy transition towards a more carbon-neutral, fossil-fuel-free society. Moving towards more sustainable energy harvesting and storage technologies that become part of a circular economy is essential. Materials sustainability will become an overriding factor in the years to come, conjointly with consumer demands that are ever-increasing. Under such a scenario, the production of Li-ion batteries is expanding considerably across the globe reviving the issue of finite Li[1] reserves, and the volume of critical raw materials such as cobalt and nickel, as pertinent examples. This concern has driven researchers to explore new, potentially more sustainable chemistries, including Na-ion,[2,3] metal–air chemistries Li(Na)–$O_2$,[4,5] Li–S,[6,7] and multivalent (Mg, Ca)[8] redox flow batteries (RFBs) and aqueous-based technologies.[9] Electrode degradation represents one important challenge to the development of rechargeable batteries. While the existence of this challenge is well known and widely researched, the electrochemical processes responsible are continually being examined and assessed.[10-14] Unsurprisingly, attempts to understand these processes have involved analysis of the composite materials. Data of this nature has vastly improved the performance of batteries and the understanding of degradation processes. However, such research has been limited mainly to the study of electrode materials in isolation, i.e., before or after cycling, failing to capture short-lived or metastable processes. There is no surprise that a further understanding of battery materials requires their analysis during operation. These requirements call for state-of-the-art methodologies. Since 2008,[15] many studies have adopted such an approach with great success. However, as it becomes a necessity for batteries to be charged every day, the need for more power has grown at a rate far greater than the improvements made in battery capacity.[16]

New analytical methods and lab-scale capabilities for existing large-infrastructure techniques are needed to probe specific battery chemistries and speed up analysis, screening and prediction of behaviour, performance, and safety for future electrification, so that new technologies can be brought to the market more rapidly to meet societal demands. Advances could include the development of remote, non-invasive and passive *operando* techniques[17] to complement present battery management systems[18] – but a real impact would come from powerful methods that are accessible outside synchrotron facilities, and operable in any lab. Right now, the consensus for small and EV-level batteries is that failures modes need to be tracked in real-time, and as such many battery companies are investing in accessories for advanced operando methods for their new battery chemistries. A recent commentary in Ref.[19] deals with exactly this need.

Rather than just probing a single material, operando techniques have provided an understanding of how the various materials function cooperatively, particularly within the working battery. This has been particularly important for the understanding of the electrode/electrolyte interface. Operando imaging and spectroscopic techniques such as NMR[20], MRI[21] and electron paramagnetic resonance (EPR)[22] have recently enabled the visualization and quantification of Li (and Na) microstructures. While these measurements do not simultaneously address phase, structural changes, and material arrangement during charge-discharge, they nevertheless offer powerful insight into side reactions and surface chemistries of battery materials. Recently, Co addressed the efficiency in charge passed in Sn and other electrode materials to address the issue of side reactions using operando NMR and neutron depth profiling to quantify the amount of alloyed or intercalated Li as a function of charge[23-25], directly imaging the coulombic efficiency of the overall convoluted processes in the anode material. Methods for imaging the movement of reaction fronts and reaction kinetics are becoming increasingly powerful. In situ high-resolution TEM[26,27] has been at the forefront of these advances, providing in-depth insights into interfacial processes and Li-driven local structural changes (for example in $SnO_2$ and Si) with a much higher resolution than MRI methods. Significant advances in synchrotron methodologies have allowed spectroscopic, two-dimensional transmission X-ray microscopy, full three-dimensional tomographic reconstructions and diffraction-based methods to be performed (such as Bragg coherent diffraction imaging), often in parallel; these have been used, for example, to track particle expansion with lithiation[28], and the evolution of strains and movement of dislocations within particles[29]. If materials exhibit magnetic character, it is also feasible to monitor lithiation-induced changes to magnetic properties as a sensitive probe of charge-discharge dynamics in a few materials, and supercapacitive behaviour in hybrid materials with ferromagnetic character.[30,31] Optical methods that probe index changes in materials that undergo intercalation/alloying vs pseudocapacitive charge storage and compensation mechanisms are also feasible in principle.

Revolutionary insights into the phenomena that underlie the operation of energy storage devices has the potential to uncover not only the basis for material selection, but how material at electrode level, and arrangement, affect performance. It may generally provide a way of analysing open questions and performance limitations including degradation mechanisms in a whole range of technologies including electrochromics, supercapacitors, solar cells, photoelectrochemical systems or water splitting where light-matter interactions are essential. One important consideration in all analytical techniques, some of which are addressed in this review, is the importance of expertise in operation, processing, and interpretation of the data. A recent report on atom-probe tomographic analysis for Li-ion battery materials[32] is a recommended read, where it becomes clear how many parameters that are technique-specific need to be considered for any meaningful analysis of battery materials by such advanced, deeply-probing and high resolution methods. Artefacts abound, and expertise in instrumentation, acquisition, processing, and interpretation is critically important.

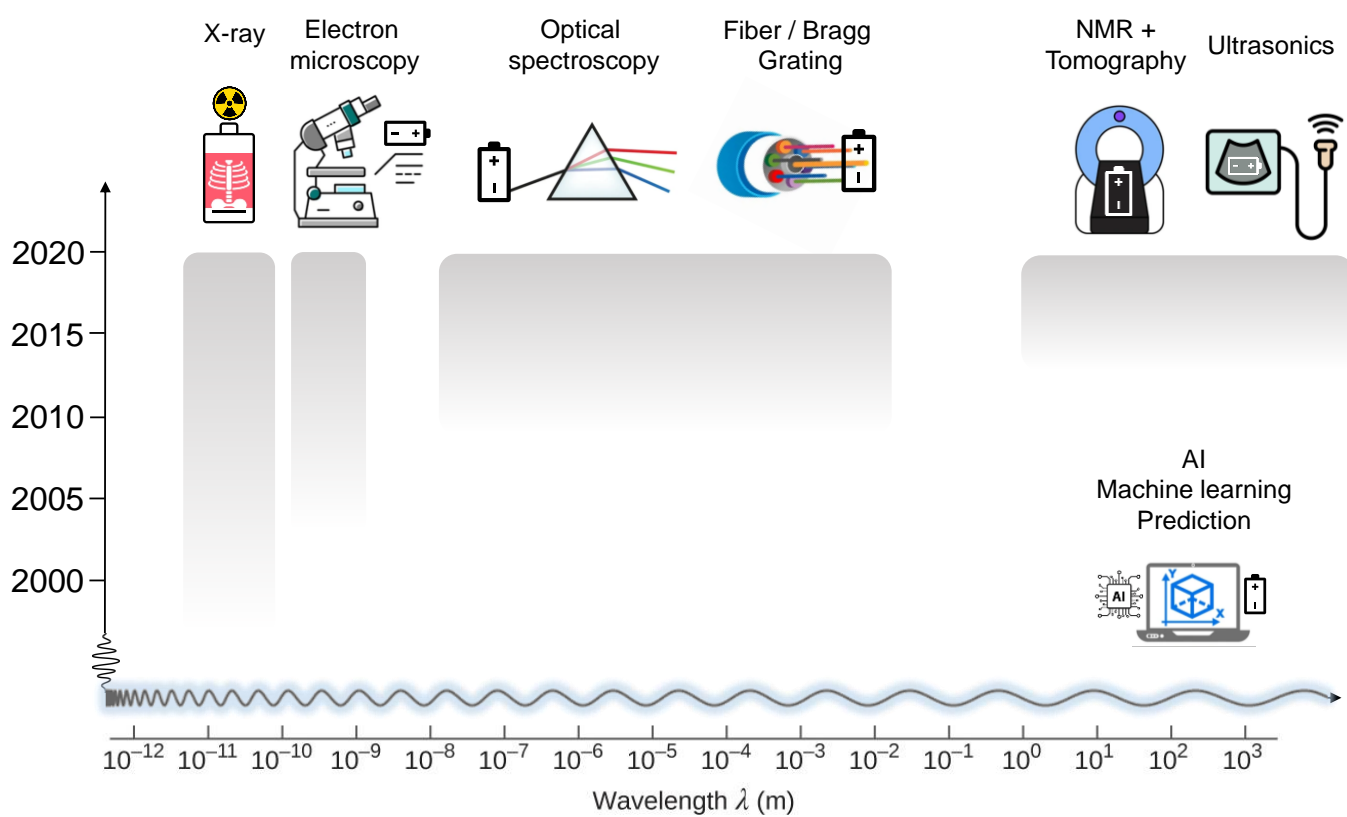

**Figure 1.** Advancement of operando non-destructive method for examining batteries and battery materials in the last two decades. Specific references for many of these methods can be found throughout this review. Notably, method span the breadth of the electromagnetic spectrum and since 2020, many of these technique can be done in synchrontron facilities and in laboratory settings with benchtop-sized infrastructure. AI, Machine learning and computational prediction underpins many modern advancement to speed up discovery and troubleshoot potential issues prior to large scale financial and time-investment.

Figure 1 depicts some of the operando methods that have recently been developed over the last two decades and the method span the energy spectrum from X-rays through to very low frequency radio waves method. Many new method are being developed specifically for batteries, but the need to analyse important materials, interfaces and electrochemical reactions seeded the research that led to new operando cells, synchrotron techniques and analytical method assisted by powerful computational approaches, to smaller benchtop/laboratory system that are also very powerful in their capability. The terms ex-situ, in-situ and operando are used throughout this work to describe measurements of battery materials. Ex-situ measurements involve cycling a cell, stopping the process at a specific state of charge (SOC), and extracting the electrode before measurement. For in-situ measurements, data collection occurs without extracting the electrode. For operando measurements, data collection is performed during electrochemical cycling. Operando techniques can be considered a special case of in-situ, with the two terms often being used interchangeably throughout the literature. In this review, the distinct differences between the two modalities are distinguished with detail on the usefulness of operando techniques.[15,33]

The objective of this work is first to provide an overview of some of the most powerful non-destructive methods used to examine battery electrode materials in real-time to date. Secondly, we will summarize some of the most promising initiatives being developed to accurately measured materials, cells, performance, and diagnostic analyses related to battery performance and health assessment in real-time. For each non-destructive method described here, the progression towards useful operando measurement and understanding, and its combination with complementary analytical techniques to monitor the electrochemical processes occurring in electrochemical cells, is of vital importance to the development of the next generation of batteries.

## 2. In situ/Operando X-Ray Methods for Battery Electrode Materials

X-ray diffraction (XRD) is a non-destructive analytical technique often used to obtain detailed information about the structure of crystalline materials, and the chemical and physical properties they possess. Battery electrode materials are very often crystalline, with highly ordered structures that allow control of ion and electron transport throughout. As a result, XRD is a powerful technique for characterizing these materials when conducted and analysed with due care and rigour for measurement and interpretation.

There are many variations of XRD used to characterize battery materials. In Kirshenbaum et al.,[34] energy-dispersive x-ray diffraction (EDXRD), was used to gain insight into the composition of an Li/Ag$_2$VP$_2$O$_8$ electrochemical cell. EDXRD is suitable for in-situ analysis of materials due to its rapid data collection rate and fixed scattering angle, the latter allowing for high-accuracy measurement of lattice parameters present in the material. In the study, spatially resolved in situ EDXRD was performed at two discharge rates and the non-discharged cell was combined with ex-situ x-ray absorption spectroscopy (XAS) measurements to determine the optimal conditions for a conductive network to be formed resulting in spatial resolution of the electrochemical reduction process within the electrode inside an intact electrochemical cell.

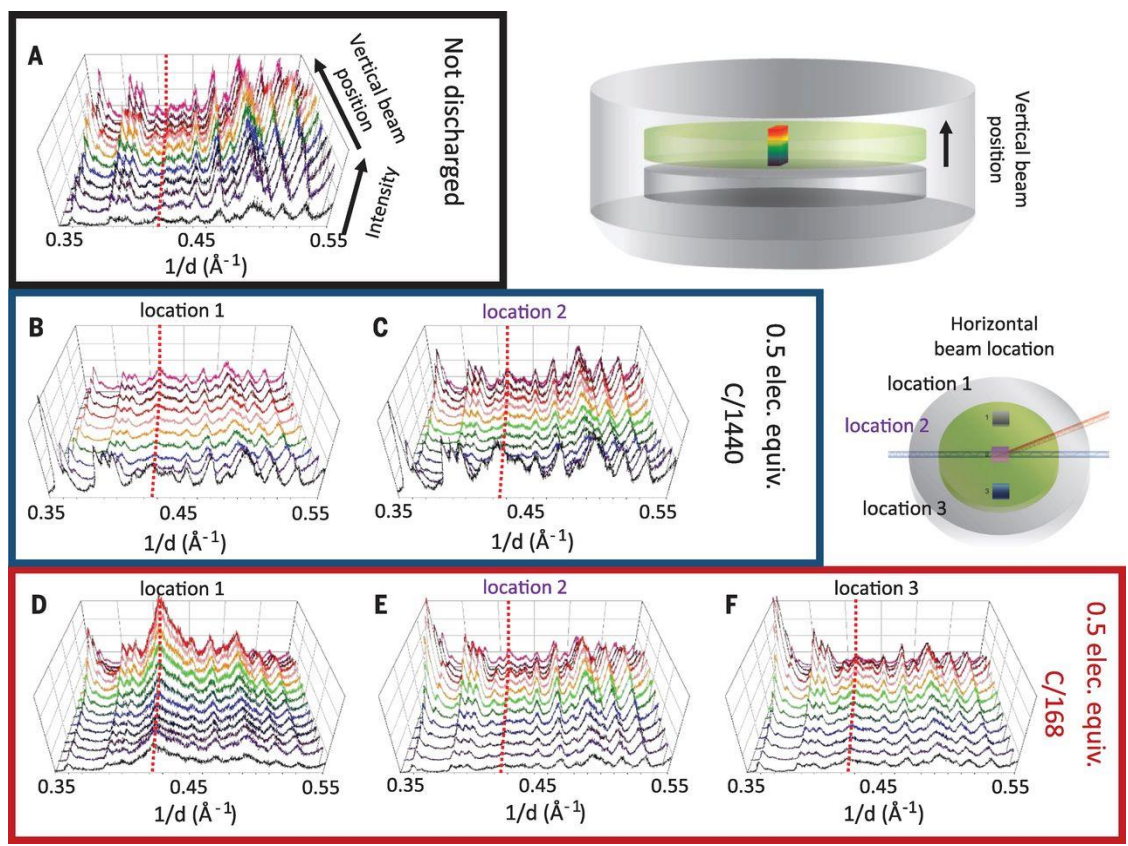

**Figure 2.** Spectra resulting from spatially resolved in situ EDXRD measurements within three coin cells: (A) A non-discharged cell, (B and C) two different locations within a cell discharged to 0.5 elec. equiv. at a rate of C/1440, and (D to F) three different locations within a cell discharged to 0.5 elec. equiv. at a rate of C/168. Spectra were obtained every 20 mm through the cathode; for clarity, only half of the scans are presented in the figure. Spectra toward the top of the figure (red hues) were obtained within the side of the cathode closer to the stainless-steel coil cell top, and spectra toward the bottom of the figure (black hues) were obtained closer to the Li anode. Adapted from ref. [35]. Copyright AAAS 2015.

Building on the in-situ study above, spatially resolved EDXRD was utilized once again to measure battery electrode performance. In the previous study, the combination of ex-situ and in-situ data provided insight into the electrochemical processes at different states of charge. Here, an in operando study was performed with measurements obtained during cycling.[36] Consequently, short-lived, metastable phases could be monitored which were not previously accessible. One of the most important aspects of the data here is the ability to track structure formation within the cell materials. The EDXRD data provided insight into the locations at which different phases occur, providing a tomographic-like profile of the cell as shown in Figure 2.

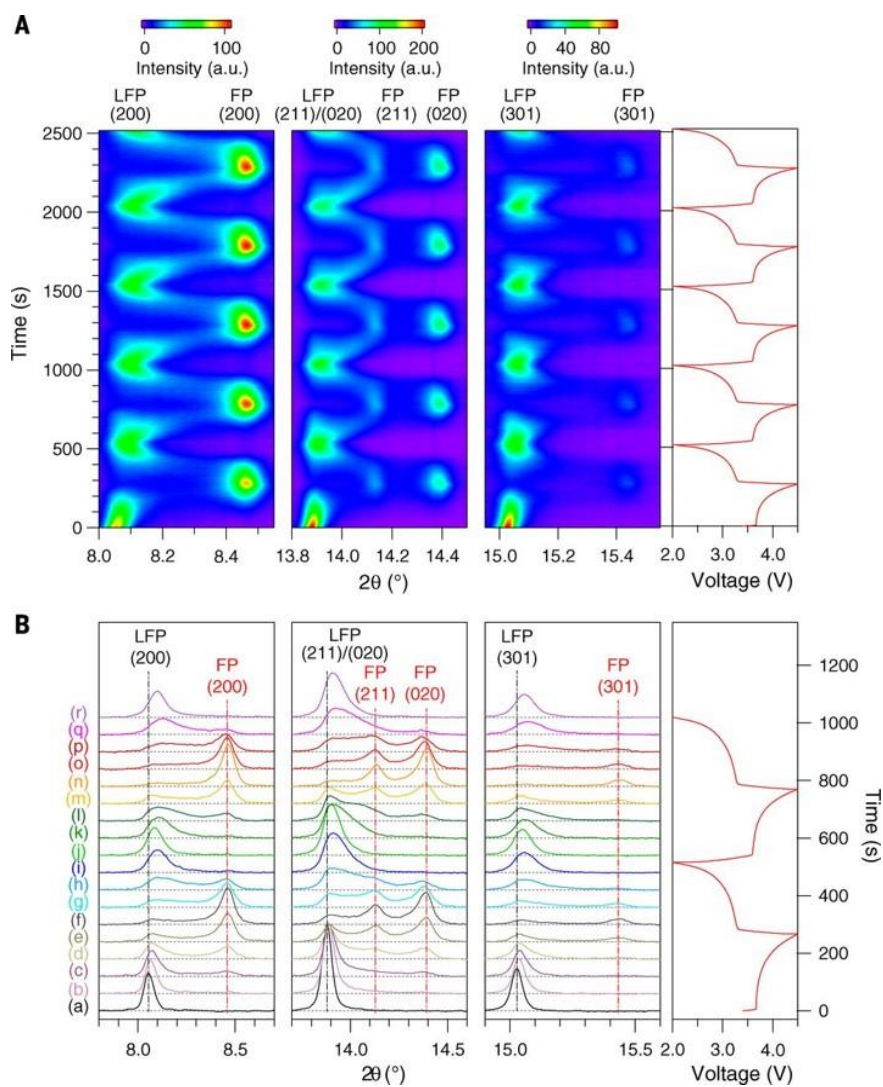

**Figure 3.** *In situ* XRD patterns during galvanostatic charge and discharge at a rate of 10 C. (A) Diffraction patterns for (200), (211), (020), and (301) reflections during the first five charge-discharge cycles. The horizontal axis represents the selected 2θ regions, and time is on the vertical axis. Diffraction intensity is colour coded with the scale bar shown on top. The corresponding voltage curve is plotted to the right. LFP, LiFePO$_4$; FP, FePO$_4$. (B) Selected individual diffraction patterns during the first two cycles and voltage profile. The baseline is represented by horizontal dashed grey lines. Black vertical lines mark the positions of LiFePO$_4$ peaks at the start of reaction; red vertical lines mark the position of FePO$_4$ peaks formed during the first cycle. Adapted from ref. [37]. Copyright AAAS 2014.

Using this data combined with impedance measurements, changes in battery resistance can be analysed. Study of both the structural changes and impedance in the cell (electrochemical impedance spectroscopy (EIS)) simultaneously during cycling enables rationalization of the changes in battery resistance, which determines the nature of the discharge mechanism. Compilation of these results provides an overview of the conductive pathways formed from reduction products and the impact on the discharge mechanism. The two techniques prove complementary and the relationships between cathode reduction, electrochemical cell discharge and impedance are established.[38] Insights into the electrochemical and structural processes which can occur in batteries are gained here as a direct result of the use of XRD methods. These are the determination of the conditions required for an optimal conductive network in the cathode and the elucidation of a rate-dependent discharge mechanism. The method used in this study demonstrates the promise of the use of in-situ XRD in the examination of battery materials and outlines the advantages of combining the data obtained from ex-situ and in-situ techniques.

As discussed in Kirshenbaum et al. above,[39] in-situ XRD provides the ability to detect certain phases which occur at different SOCs of a battery including material phases that are difficult or often impossible to be reliably detected ex-situ in a post-mortem analysis from cell-disassembly. However, the lifetimes of certain processes are smaller than the time taken to disassemble the cell and perform the required measurements and others are extremely sensitive to changes in their local environment, which can include access to changes in electrolyte, salt concentration (even when carefully contained in battery-grade storage environments) and especially to water and ambient air for more typical post-mortem analyses. Therefore, operando techniques must be used to capture this information. In Liu et al.,[39] operando XRD with high temporal resolution was used to analyse metastable structures formed during the high-rate cycling of $LiFePO_4$ electrodes and determine the effects of strain vs compositional variation. These results are compared to simulated XRD patterns of some of the structures predicted to form during cycling. Disagreement between data indicates the formation of metastable structures not detected ex-situ, i.e., particles with lattice parameters that vary from the predicted composition. The results are also used to quantify the compositional variation during cycling. The operando XRD patterns (Figure 3) show the absence of any particle interface (disappearance of Bragg reflections corresponding to second particle), indicating that strain effects did not occur. Based on work prior to this study, reactions in the electrode material were predicted to proceed via a process which would cause, using conventional wisdom, poor battery performance, while measurements indicated the contrary. The utilization of operando XRD here provides the ability to explain this disagreement, by capturing processes which could not be detected previously using ex-situ or even in-situ methods.

In general, X-ray methods used to probe battery electrode materials involve high energy penetration of the physical cell. However, in Lu et al., a novel X-ray nano-computed tomography (CT) dual-scan superimposition (DSS) technique is used to probe some of the key parameters of electrode materials which affect battery performance.[40] A 3D volume of a Li-ion half-cell is reconstructed with an NMC111 cathode vs a Li metal anode, and achieved at the nanoscale for the first time. The study investigates how the structural characteristics of the cathode such as porosity, tortuosity and thickness affect electrochemical processes within the cell such as SoL, electrolyte salt concentration ($C_{EY}$), Li-ion flux and charge transfer at different rates. Adaptations to the initial microstructure were also made to probe the evolution of the same electrochemical processes in novel microstructure designs.

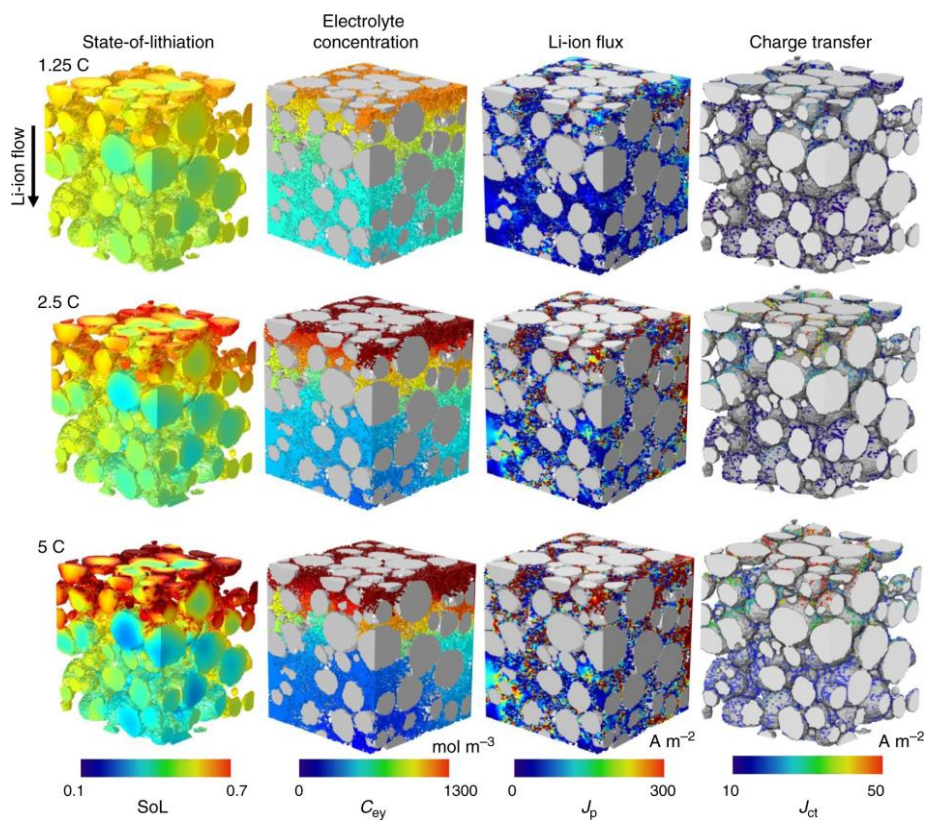

**Figure 4.** Simulated discharge of the reconstructed cathode and an ideal lithium anode (half-cell) at 1.25, 2.5 and 5 C. The ionic current flows from the top (separator side) to the bottom (current collector side). Field variables at 50% depth-of-discharge (DoD) are shown here. Reproduced from ref. [40]. Copyright Nature Publishing Group 2020.

Figure 4 shows the computed evolution of the half-cell during discharge at three different rates of 1.25C, 2.5C and 5C, from top row to bottom row. Each individual schematic depicts the half cell with current flow in the through-thickness direction. The top corresponds to the region of the cathode closest to the separator. The bottom represents the region closest to the current collector. Each column represents one of the key electrochemical parameters of the cell.

In the first two columns of Figure 4, SoL, and electrolyte salt concentration of the cell at 50 % depth of discharge (DoD) is shown. At the lowest rate, there is a uniform distribution observed in each case. As the rate is increased, a severe gradient develops due to the competition between mass transport dynamics and reaction kinetics. Lithiation occurs only where the active material, NMC particles in this case, are exposed to the Li ions flowing through the electrolyte in the form of Li salts. As the cell is discharged at low rates, the Li ions flow along the electrode slowly, maximising the amount of active material they access. As the rate is increased, Li ions bombard the active material closest to the separator, maximising charge transfer in this region. Consequently, mass transport is slow, meaning that the ions are unable to supply reactant into deeper regions of the electrode at a sufficient rate. NMC close to the current collector are therefore not utilised, resulting in severe capacity loss in the cell.

The third column shows the distribution of Li$^+$ ion flux $J_P$. At low C-rates (1.25 C), the flux is evenly distributed. At higher C-rates (above 2.5 C), heterogeneity is evident due to the non-uniformity of the pore microstructures, with much higher flux in narrow pores. High flux causes local Ohmic heating, which can lead to electrolyte decomposition and thermal runaway. The fourth column shows the charge transfer $J_{CT}$ which represents reactivity. The reactivity distribution is shown on a particle-by-particle basis, with reactivity more homogeneous at low C-rates. $J_{CT}$ is much higher close to the separator than near the current collector, due to the mass transport limitation already mentioned.

The simulations based on the reconstructed cathode microstructure show that the inhomogeneous lithiation that occurs during cycling occurs primarily due to the broad distribution of particle size. As a result, changes in both the level of porosity and pore size in the direction of lithiation leads to a variation in lithium-ion diffusion path lengths. The imbalanced utilization of NMC lowers the energy density of the cell, which inhibits the long-term structural integrity of the electrode. Subsequent simulations were performed to investigate the use of next generation electrode designs, exploring the effects from the manipulation of particle size, pore size distribution, and electrode compression, to probe the effects of specific structural alterations to the performance of the cell.

A full interrogation of the causes of inhomogeneous lithiation based on the microstructural size distributions of the electrode and how this leads to poor battery performance is provided. The gradients of the electrolyte concentration and SoL increase with C-rate and depth of discharge, highlighting the issues that occur in battery performance at higher rates. It is particularly evident that particle size, pore size and the level of porosity are the main factors which determine the distribution and direction of flow of lithium ions during discharge. Uneven intercalation between particles occurs as a result, leading to the underutilization of capacity and reduced power density. The study also describes the effects of calendaring on the microstructural evolution of the electrodes, a further look at the process can be found elsewhere.[41] Overall, the DSS computational technique and modelling used here represents a truly non-destructive method for investigating electrode degradation mechanisms based on the distribution of electrode particles within the cell.

More recently, X-ray computed tomography has moved to the lab and analysis of battery materials' absorption coefficient is possible with voxel sizes down to ~50 nm. The most recent approaches combine artificial intelligence and analysis techniques including machine learning, with nano-CT on the lab scale. The motivation has been to develop powerful and accurate predictive models of the electrochemical response of battery materials and electrodes, from variations to microstructure and composition variations that can be visualised in 3D with post mortem analysis and predictive models for new materials or existing materials under various operando conditions. X-ray CT has superior resolution compared to nano-CT imaging, but developments in high-resolution nano-CT systems are underway. The technique is a non-destructive tool and provides opportunities for operando investigations in battery materials and solid-state battery interfaces and solid electrolytes, to name a few. Combing nano-CT imaging data and computational modelling, reconstructed volumes, and parameters such as pore tortuosity, particle size etc. can be extracted and incorporated into the models to simulate electrochemical response. Machine learning and related approaches are necessary to obtain accurate and representative volume of the material under examination in CT images. Some recent reviews have outlined the details and opportunities of X-ray CT[42,43] and also computationally-supported nano CT imaging[42] for battery research. The importance of data quality, interpretation and curation is a consistent in the application of CT imaging for battery materials analysis and extracting meaningful data requires careful workflow and data processing. Opportunities exist not only for 3D visualization but for obtaining digital twins of battery electrodes to watch their behaviour in real-time and translate this knowledge to prediction of similar material sunder different conditions, or for screening new materials of battery chemistries with particle size and interface-level resolution in real time in 3D.

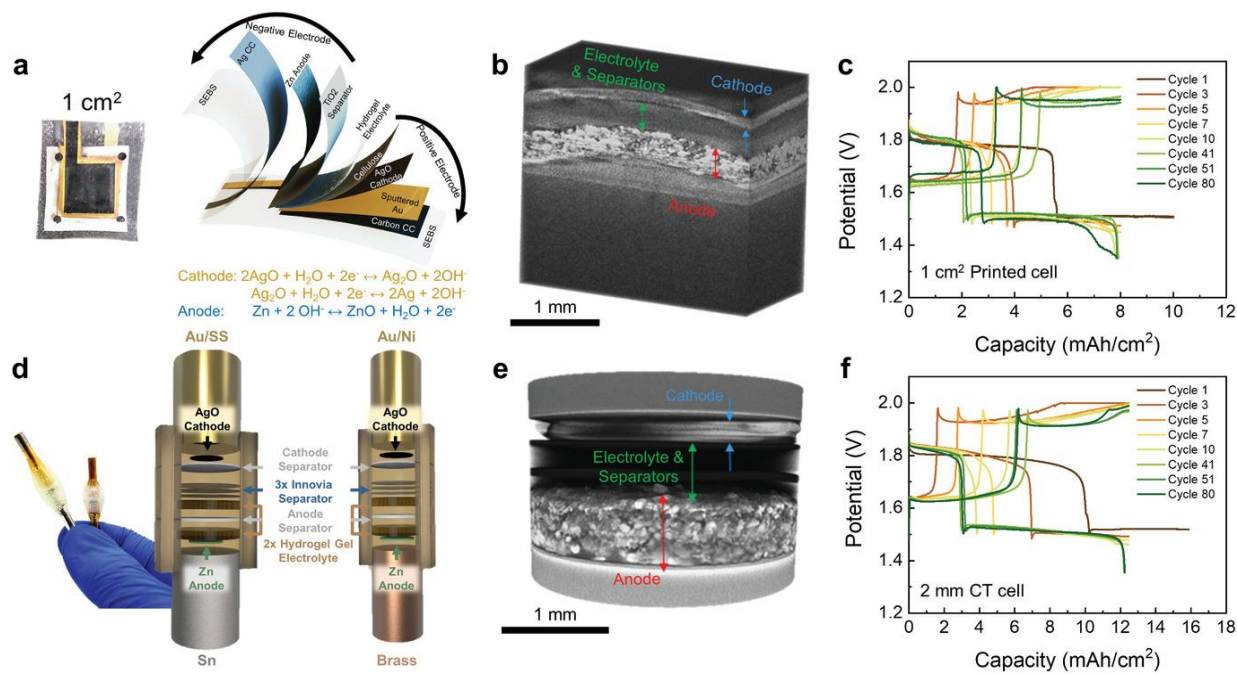

**Figure 5.** Cell design and performance of a Zn-AgO printed cell (a) and in-situ CT cell (b). 3D X-ray CT tomographic reconstructions (b, e), and electrochemical performance (c, f) for Zn-AgO printed cell (top row) and in-situ CT cell (bottom row). Adapted from Ref. [43]. Copyright Wiley VCH 2017.

As a representative example, Scharf et al.[43], showed how in-situ X-ray micro-CT in combination with electrochemical battery cell characterisation could be used to interrogate how gassing at the current collector influence the performances of Zn-AgO batteries. Their results and process, representatively overviewed in Figure 5, revealed that OER and HER gas evolution depended on the choice of current collector material. While the likelihood of gas evolution processes could in principle be extracted from basic electrochemical knowledge of this metal in such an electrolyte, the micro-CT provided useful additional quantification on z-axis volumetric expansion within the electrode. Simply changing the current collector made an obvious improvement in performance giving a high cycling capacity of 12.5 mAh cm$^{-2}$ for more than 325 cycles in a 4 cm$^2$ form factor, corresponding to volumetric energy density of 100.6 Wh L$^{-1}$. The development here of the in-situ cells is also important for future application of micro-CT and related methods, as the cell construction is a basic necessity to allow accurate probing by CT instrumentation and they showed how the approach could provide meaningful examination of cell degradation over long term cycling.

In-situ or operando X-ray techniques represent a powerful array of tools which can be used to investigate the electrochemical behaviour of battery electrode materials at different states of charge or discharge/during cycling. As most battery materials are crystalline, their examination using X-ray methods provides information on changes to their structure which can be correlated to electrochemical processes and overall battery performance. The major disadvantage of this family of tools is their inability to detect amorphous phases, such as electrolyte species and the SEI layer. Involving computational methods to enable digital twins with accurate processing of the material volumes being imaged is essential so that examination of representative cells can provide predictive capabilities for new materials and batteries.

3. In-situ/Operando Nuclear Magnetic Resonance Analyses of Battery Materials

The performance of battery electrodes relies on the behaviour of electrons and ions within and in the vicinity of the material. Nuclear magnetic resonance (NMR) spectroscopy characterizes the materials based on the electron density present by applying an external magnetic field of strength $B_0$ to measure the frequency at which atoms in the material achieve resonance. The resonant frequency of the nucleus relative to the applied field is known as the chemical shift, usually measured in parts per million (ppm). Electrons orbiting the nucleus naturally oppose $B_0$, referred to as shielding the nucleus, which determines the chemical shift. The method is useful for atoms with nonzero nuclear spin. External interactions refer to those which involve the nuclear spins (rotational movement of nucleus around its axis) and the magnetic field. Internal interactions determine the NMR spectra, dictating signal shift and line shape. These internal interactions are described in detail elsewhere.[39] NMR itself is not phase sensitive. Therefore, to determine the structure of battery materials, NMR spectra must be used with line shape analysis and known shift ranges.

NMR offers insights into particle sizes, crystal structure, surface changes, phase analysis, oxidation states of elements and electrochemical performance of battery materials. While XRD provides similar information for crystalline materials, NMR proves useful for amorphous materials also.[37] Given batteries consist of complex material systems under continuous modification in charging and discharging, predictions of the properties of battery materials using theoretical approaches such as density functional theory[44] and Hartree Fock methods have been effective and useful predictors in concert with NMR experiments.[45,46] The combination of NMR and diffraction methods can be extremely powerful, as the techniques provide short-range, local structure information and long-range information, respectively. Insight into crystal and electronic structure can be provided simultaneously. NMR proves useful regardless of the charge storage mechanism, given that alloying, conversion, and intercalation mode cathodes all involve significant structural changes and phase transformations during cycling.

Real-time, non-destructive NMR measurements are required to adequately assess the evolution of the battery during electrochemical cycling. For in-situ NMR measurements, self-relaxation processes are minimized, high chemical specificity can be achieved for both crystalline and amorphous materials, and both short-lived and metastable phases can be tracked. Ex-situ NMR does not provide the necessary information to understand the behaviour of the battery during cycling. However, ex-situ NMR can be used to streamline the in-situ experiments by selecting the relevant NMR parameters to study for well characterized materials. The in-situ approach has its limitations, it is performed under static conditions such that sample spinning is not an option. Therefore, cell design, hardware, detection, and analysis methods must be given far more thought.

In Key et al.,[47] a range of NMR techniques used to probe local structural changes during discharge of crystalline silicon in a working LIB are described. Figure 6 shows the stacked in-situ spectra obtained, at a range of capacity levels, revealing the change in composition of the electrodes at each SoC. The work outlines the benefits of using the combination of NMR and diffraction methods as a diagnostic tool for battery electrodes. Similarly, the advantages of using the combination of in-situ and ex-situ analysis are shown, along with the accompanying complications that arise in the setup of the required system. The data obtained provides the ability to streamline in-situ NMR measurements such that comparison between the ex-situ and in-situ data can be achieved. The in-situ NMR measurements detect phases which are not observed in the ex-situ data, a clear indication that amorphous phases form during the cycling of Si electrodes in LIBs. Dynamic processes which are short lived or metastable and therefore not accessible via ex-situ methods are captured. While analysis can prove difficult due to the extremely short time scale of the interactions involved, NMR is well suited to nanosecond to sub-nanosecond time domains.

The key result here is that the combination of in-situ and ex-situ NMR with a knowledge of the structural and electrochemical processes at work provides a clear route to investigating the processes that occur in the vicinity of the electrodes in an LIB, especially processes related to the formation and passivation of the SEI layer. The requirement for a more sophisticated methodology design for in-situ measurements compared to traditional NMR measurements is evident, including the inclusion of a flexible electrochemical cell to facilitate the in-situ probe.

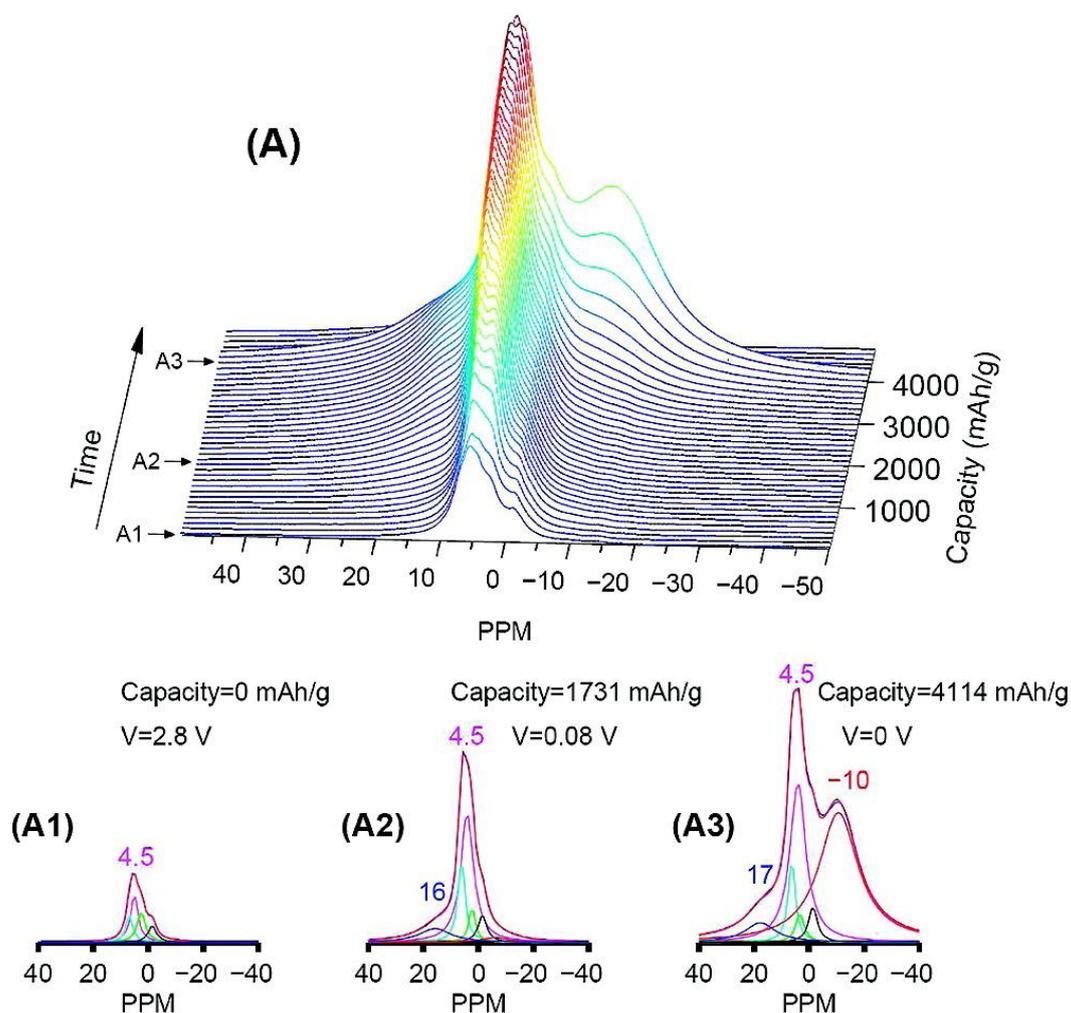

**Figure 6.** (A) Stacked plots of *in situ* $^7$Li static NMR spectra of the first discharge of an actual crystalline Si vs Li/Li+ battery (the colour bar shows the relative intensity scale for the spectra). (A1-A3) Deconvoluted spectra at various discharge capacity values of interest. Adapted from ref. [15]. Copyright ACS 2009.

In applying NMR to battery materials, the interfaces between the electrodes and the species with which they react (e.g., electrolyte, SEI layer) have a large impact on performance. Decomposition products are one of the primary factors which dictate ion transport through interfaces and the electrolyte, ultimately determining energy density. Interference in NMR spectra is often caused by dendrite formation, which can lead to short circuiting and battery failure. Efforts to combat these effects in various battery systems are widely researched.[48-51] This phenomenon occurs in lithium and sodium cells with metallic anodes during cycling. One of the limitations of NMR is that the RF fields used to excite nuclear transitions is severely limited through metal samples because of an effect known as skin depth. Skin depth is usually an order of magnitude larger than dendrite thickness. Therefore, signals penetrate dendrites fully but appear constant relative to the signal of the bulk metal.[52] Identification of microstructural characteristics, growth mechanisms and contributory factors is required to minimize this interference during in-situ NMR experiments to provide a better insight into the evolution of the microstructures leading to electrode degradation.

Lithium metal batteries (LMBs) are being investigated with renewed interest for higher energy density future battery systems.[53-55] Lithium metal is well known to have the highest theoretical specific capacity of all lithium-ion anodes and low negative potential. In LMBs, charging and discharging involves the deposition and stripping of Li metal for each process, respectively. Capacity retention, cycling stability and fire hazards are the key issues for in service LMBs, which can all be attributed to Li dendrite growth, but many examples of these batteries are deemed safe. Li dendrite growth and the factors that affect it for LMBs being investigated in new form factor cells or at non-standard temperature, or at higher charging rates can be difficult to accurately track and examine in real-time during battery operation. Figure 7 shows the stages of Li dendrite growth in tandem with SEI formation, the predominant mechanism leading to short circuiting in LMBs.

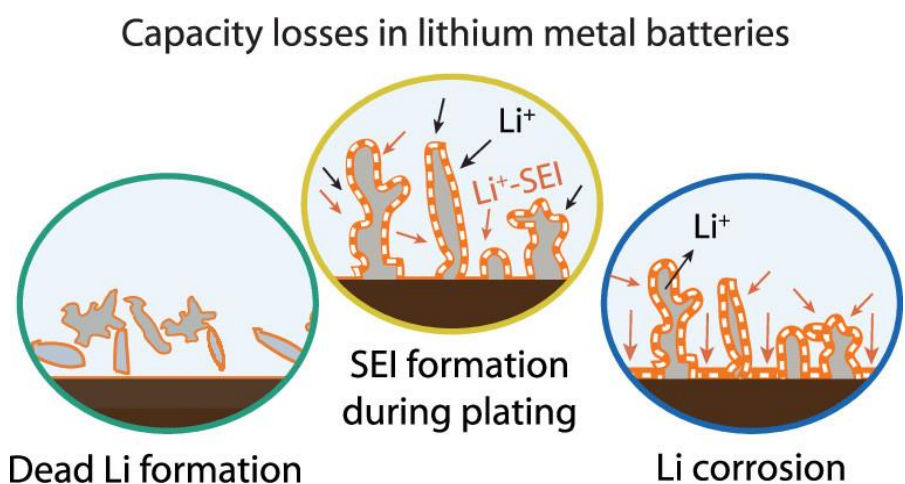

**Figure 7.** Lithium dendrite growth: Schematic depicting the SEI build up and dead Li formation at the electrode/electrolyte interface. Reproduced from ref. [56]. Copyright ACS 2020.

Gunnarsdóttir et al. demonstrate an in-situ NMR metrology where an "anode-free" LMB setup is employed.[56] The typical Li metal anode is absent, meaning that Li is plated directly onto a bare copper current collector from a $LiFePO_4$ cathode. The build-up of dead Li and the evolution of the SEI layer can then be tracked by in-situ NMR. While traditional performance analysis techniques detect only the capacity losses of the system, in-situ NMR can be used to deconvolute individual capacity drops to pinpoint their sources.

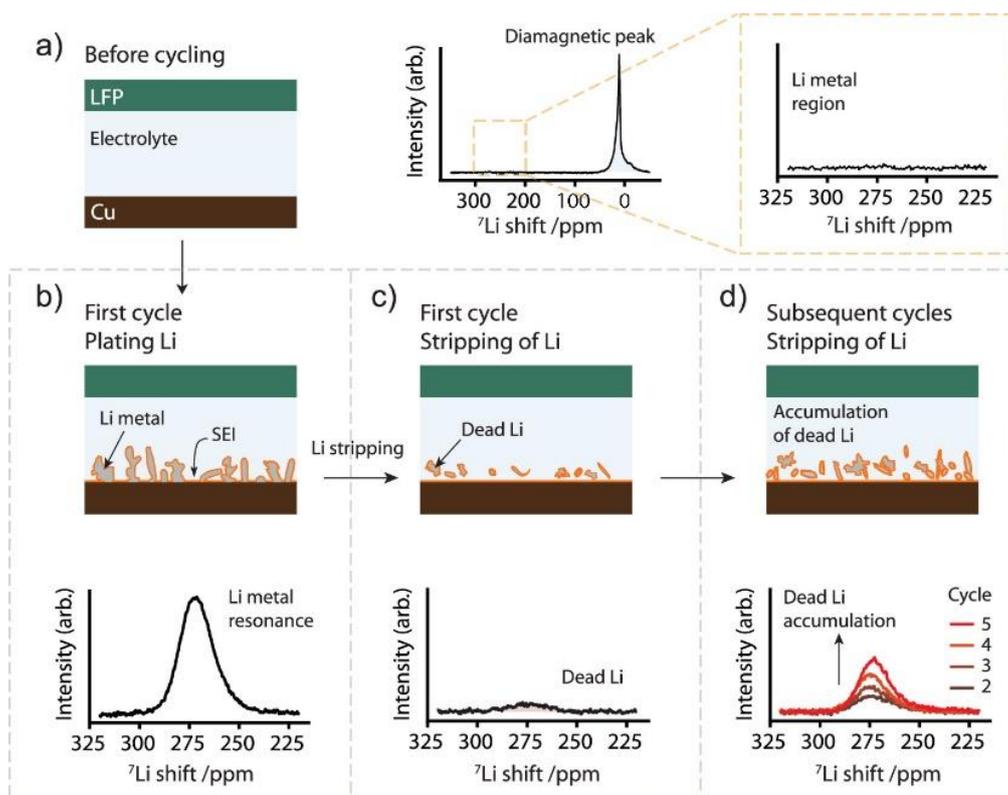

**Figure 8.** Schematic of the $^7$Li *in situ* NMR technique used to study dead Li formation and the resulting $^7$Li NMR spectra. (a) The Cu-LiFePO$_4$ (LFP) cell before cycling and the corresponding $^7$Li NMR spectrum showing the resonance of the diamagnetic Li (the SEI and Li+ ions) and the absence of the Li metal peak. (b) Charging the cell results in Li deposition, as shown in the $^7$Li NMR spectrum of the Li metal region. (c) At the end of discharge, the Li metal signal can still be observed, which is attributed to dead Li. (d) Further cycling of the Cu−LFP cell results in an accumulation of dead Li over the next cycles (cycle 2−5), the intensity of the Li metal signal increasing at the end of stripping in each cycle. Adapted from ref. [56]. Copyright ACS 2020.

In-situ NMR can detect the BMS shifts caused by Li metal and determine the location and environment of the sources of capacity losses by providing information on the density of surface coverage of the current collector in its various environments. In this study, the environment is varied using different electrolytes and coatings. SEI formation is of paramount importance during cycling of a battery, yet its evolution can either be key in optimizing performance or detrimental to it. By monitoring the surface coverage of the current collector, the effects of the SEI layer on performance can be determined. Figure 8 (a) shows the Cu-LiFePO$_4$ cell before cycling, and the NMR spectrum shows the diamagnetic peak corresponding to the SEI and Li+ ions, with the spectrum region associated with Li metal highlighted to show its absence. The schematics in Figure 8 (b-d) show the cell at different stages of cycling, providing an image of the accumulation of dead Li over time. The NMR spectra associated with each stage are also shown. For a system with no dead Li deposited, there would be an absence of any peak for NMR spectra acquired after discharge, i.e., the stripping phase. The spectra clearly demonstrate the build-up of dead Li as the number of cycles increase. A continuation of this trend would lead inevitably to contact between anode and cathode, resulting in short circuit and system failure.

The results of this study highlight the significant advantages in using in-situ NMR metrology to analyse battery degradation processes, in this case Li dendrite growth. The "anode-free" cell removes some of the complications of NMR methods for battery cells, like the issue of spacing between electrodes within the cell and effectively locating Li metal build-up. Potential solutions are also examined in the form of metal coatings, additives, and artificial SEI. Once again, in-situ NMR applied to the "anode-free" setup can be used to track the dead Li formation for each environment. Going forward, the versatility of this approach is evident. Small alterations can allow its extension to Li-S and Na metal batteries and operando studies. The fact that disassembly of the cell is not required means that the approach in this study can be added to the arsenal of non-destructive techniques for monitoring battery electrode degradation mechanisms in real time.

The geometry and composition of commercially used battery cells render many direct NMR methods ineffective in examining battery performance. The spacing between electrode layers and conductive casings prevent the required radiofrequency penetration. In Pigliapochi *et al.*,[57] an ultrafast, inside-out NMR technique is used for the assessment of rechargeable cells. Here, cell classification is performed via a single shot free induction decay (FID) NMR spectrum acquisition of the liquid compartments surrounding the cell. Changes in magnetic susceptibilities occur due to changing oxidation states in the electrode materials during charge and discharge. The method provides a non-invasive cell classification based on cathode material, volume of electrolyte and cycle life, while also presenting an overview of the ageing model of the system.

Much like the in-situ NMR techniques described thus far, a revised cell design is required to provide access. A schematic of the setup is shown in Figure 9 (a). The system is simply placed inside an NMR probe with its z axis aligned with the magnetic field. The cell was probed using inside-out magnetic resonance imaging (ioMRI), described in detail elsewhere,[58] prior to NMR measurements. However, there are limitations to the ioMRI approach. As strongly magnetic materials, like those used as battery cathode materials, cause distortions in the magnetic imaging. The limitations of the ioMRI technique can be avoided using FID NMR spectroscopy.[59] Instead of obtaining a series of images, spectra are acquired in one fast, single-shot readout lasting less than a second. The speed of the technique allows access to the fast relaxation regime, negating the effects of large magnetic field gradients. Figure 9 (d) shows the $^1$H NMR spectra obtained at 2.5V for the three cells of different cathode materials. The $^1$H NMR spectra acquired at successive states of charge in one of the cells is shown in Figure 9 (e). The change in chemical shift and signal intensity due to lithiation during the charging cycle is indicated by the black arrow.

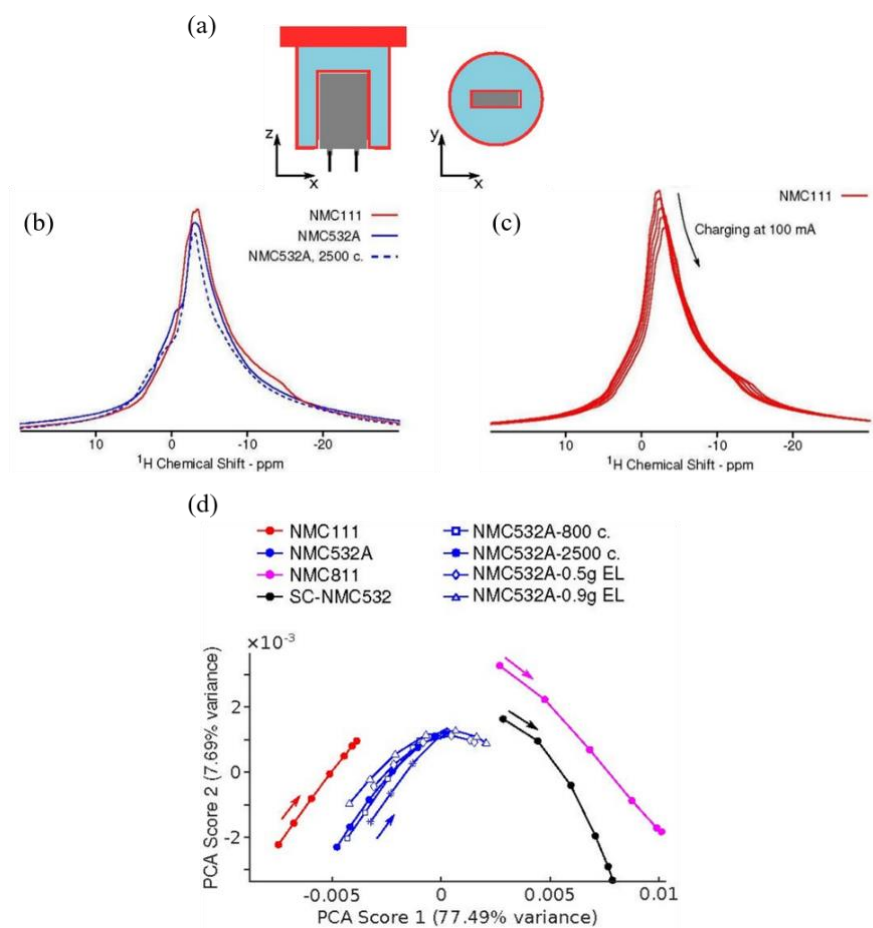

**Figure 9.** (a) Schematic of 3D-printed holder (red) used for ioMRI and NMR. The holder contains the cell (grey), which is surrounded by water compartment (light blue). The system is placed inside an NMR probe with the z-axis aligned with the static magnetic field. The vertical and horizontal cross sections are shown left and right, respectively. (b) $^1$H NMR spectra taken at 2.5 V for cells containing, as cathode: NMC111 (red), NMC532A (solid blue) and NMC532A already undergone 2500 cycles prior to the MRI (dashed blue). (c) $^1$H NMR spectra taken at intervals during charging of the cell containing NMC111 as cathode. The black arrow shows the direction of shift of the spectra as the cell undergoes charging. (d) Principal component analysis (PCA) performed on the $^1$H NMR spectra of all the cells. Each arrow indicates the direction of the trend(s) of the corresponding colour as the cell is charged. Adapted with permission from ref. [57]. Copyright Wiley 2020.

Principal component analysis (PCA) is used to classify groups of measurements based on the few strongest-contributing features of the dataset, which are the principal values, and is useful in the diagnosis of energy storage mechanisms.[60] The results from PCA applied to the $^1$H NMR spectra acquired are presented in Figure 9 (d). A clear trajectory with charging is observed for each cell. The trajectories of cells of the same cathode component and different electrolytes or cell age are grouped together (blue lines), while there is a significant difference between cells of different cathode components. The PCA analysis shows that cell classification based on cathode materials as a function of state of charge is an effective approach while electrolyte amount and cell age have little influence on the results. The ultrafast inside-out NMR technique is depicted as a fast, robust, non-destructive method for examining battery electrode materials in-situ. The versatility of the setup means that it can be extended to operando studies in the future. The applicability of in situ NMR and MRI techniques to commercial cells has proven difficult to date, calling for novel adaptations to their methodology.

## 4. In-situ Raman Scattering of Phase Changes in Battery Electrodes

Raman spectroscopy is commonly used to determine the composition of both crystalline and amorphous materials. The technique is based on the energy exchange and direction change associated with light incident on a molecule. The molecule gains energy and the energy of the photons scattered from the molecule can be measured to determine the composition of the molecule. These energies are observed as peaks on a Raman scattering spectrum, each peak corresponding to a specific molecular bond vibration or optical phonon for Raman-active molecules. The technique is particularly useful for multi-layered materials such as the porous metal oxides often used in battery electrodes. Information can be revealed on crystal structure, electronic structure, lattice vibrations and flake thickness of layered materials, and can be used to probe the strain, stability, charge transfer, stoichiometry, and stacking order.[61] The correlation between the capacity of intercalation in an electrode to the degree of disorder in the material can also be determined.[62] Such characteristics can also be analysed during cycling, once a modified in-situ cell is incorporated to permit light transmission. An optical pathway is usually provided in the form of an optical window, allowing measurements to be obtained non-destructively in real-time.[63] The window material is chosen based on its transparency relative to the frequency of incident light. Examples include sapphire, KBr, diamond and fluorite glass $CaF_2$.[64] The cells are referred to as spectroelectrochemical cells, providing the ability to perform electrochemical and optical measurements simultaneously.[65]

In Nakagawa et al. [66], the changes in the surface crystallinity of edge plane HOPG (graphite) anodes are probed using in-situ Raman spectroscopy. The corresponding spectra show that surface crystallinity is significantly lowered within only a single charge/discharge cycle. The combination of these results with considerations of the intercalation mechanism of Li$^+$ provide information on the structural degradation of the graphite surface. Conventional Raman spectroscopy is limited to near-surface regions predominantly, depending on the optical characteristics (scattering, roughness, absorption etc.) at the incident probe wavelength. More intricate adjustments are required to permit deeper penetration. Like the techniques mentioned previously, in-situ Raman measurements require a more complex electrochemical cell. Here, a three-electrode cell was equipped with an optically flat window made of Pyrex glass to permit light transmission.[62]

In Zhu et al.,[67] in-situ Raman and in-situ XRD are used to probe electrode degradation mechanisms in a lithium-sulfur battery. Lithium-sulfur (Li-S) batteries have a very high theoretical capacity of 1675 mAh g$^{-1}$. The most controversial issue related to an Li-S system is in its discharge mechanism. Here, analysis of the mechanism based on the influence of the electrolyte on the dissolution of sulfur and polysulfides. The short lifetimes of the polysulfide species mean that measurements must be obtained during cycling. In-situ Raman measurements provide the ability to probe the presence of each species and the phase changes which occur while the battery is in service. While the same species are detected in each case, the rates of polysulfide formation and diffusion to the anode depend on the electrolyte and the associated ionic mobilities of the various species in the electrolyte. In-situ Raman scattering measurements of the Li-S battery in an ionic liquid electrolyte (LiTFSI–PY13–FSI) and ether-based electrolyte (LiTFSI–DOL–DME) are performed for comparison. The presence of elemental sulfur is detected near the Li metal anode in the ether-based electrolyte at the end of charge. The results indicate that multiple reactions of the Li metal surface with soluble polysulfides and/or elemental sulfur cause the degradation of Li metal, leading to battery deterioration.

The versatile cell configurations used for Raman measurements are shown in Figure 10 (a) and (b).The former configuration allows probing of changes to the sulfur cathode during cycling, with the sulfur cathode placed in contact with the cell window.

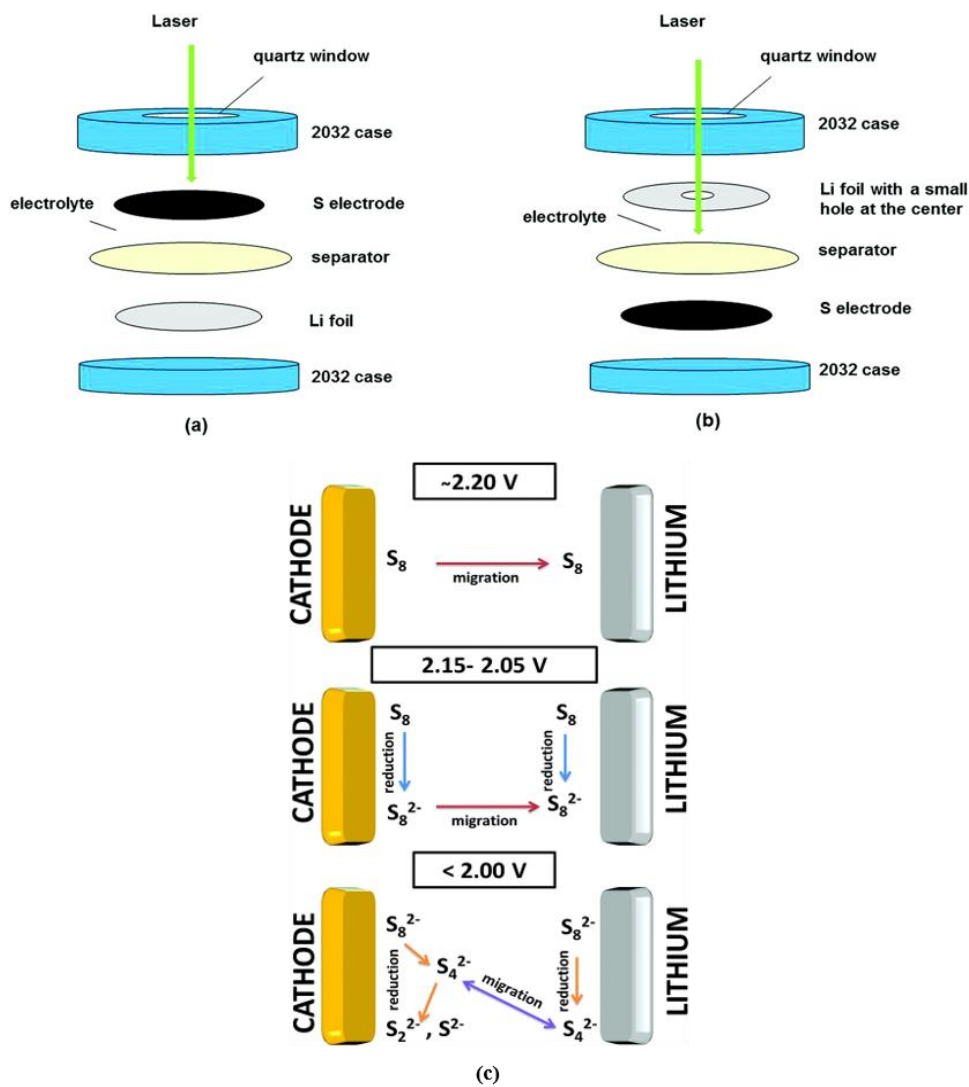

**Figure 10.** In-situ Raman cell designs for monitoring changes in (a) the cathode and (b) electrolyte near Li metal. (c) Schematic of the reduction process from *in situ* Raman measurements at both the cathode and Li metal sides depicting the evolution of polysulfide species in a Li-S battery. Adapted with permission from ref. [67]. Copyright RSC 2017.

The latter allows monitoring of the evolution of electrolyte species in the vicinity of the Li metal anode via a small hole punched in its center. The cell differs from a traditional electrochemical cell only in that the outer casing is modified with an optical window inserted to provide access to the incident laser beam. The window consists of a 0.15 mm quartz plate, transparent to frequencies in the UV-Vis and IR range commonly utilised for Raman analysis.[68] For in situ XRD measurements carried out in the study, the quartz window was replaced with beryllium. Figure 10 (c) provides a summary of the evolution of the polysulfide species occurring at both cathode and anode.

The in-situ Raman spectroscopy performed here provides a clear picture of the discharge mechanism of a Li-S battery, the effects of the chosen electrolyte and compares the phase changes of ionic species in the electrolyte in the vicinity of both the sulfur cathode and Li metal anode. The cell designs used allow measurements to be performed non-destructively in real-time, capturing the true behaviour of the system during discharge. Application to more traditional battery systems containing common electrolytes is straightforward. The Li-ion transport dynamics at particle level are revealed via in-situ Raman spectroscopy in Wei et al.[69] for a battery containing a $LiFePO_4$ cathode and electrolyte composed of $LiPF_6$/ethylene carbonate (EC) and dimethyl carbonate (DMC) through the examination of Raman peaks originating from the C-O stretching vibration in EC. The Li-ion concentration at the cathode surface was found to fluctuate based om the redistribution of $Li^+$ within the pore space of the electrode.

All-solid-state lithium-ion batteries (ASSBs) have also attracted attention due to increasingly widespread belief in their safety, relative to Li-ion batteries containing organic liquid electrolytes.[70-72] However, the major concern with these systems is the instability of the interface between the sulfide solid electrolytes and the electrodes. Once again, the electrochemical processes which govern the evolution of the interfacial layers require in situ measurements. In-situ Raman spectroscopy is used in Zhou et al. to probe degradation mechanisms at the electrode/solid interface in real-time.[73] In-situ and ex-situ Raman spectroscopy are applied in tandem to probe the evolution of sulfide species during discharge once again.

With a custom-designed cell, it is shown here that in-situ Raman measurements can be used to access the lithium/solid electrolyte interface, and the solid/solid interfaces within the layers of the composite positive electrodes. This approach represents a valuable tool in understanding the effects of cycling on the interfaces within ASSBs and the consequential routes to electrode degradation. Given that in-situ Raman can be used at a microscale level with the appropriate beam width, application to thin film ASSBs can thus clarify local structural changes in various regions within the cell. An example of such is demonstrated in Matsuda et al.,[74] where micro Raman spectroscopy is used to interrogate the local material structure at either side of the $LiCoO_2$ cathode by acquiring measurements at the front and back of the cell.

Like the other techniques described thus far, Raman spectroscopy proves most useful when used in tandem with other techniques as a complementary setup. The combination of electrochemical measurements with both ex-situ and in-situ Raman measurements can provide insight into the SoC of battery electrodes. While the superiority of in-situ Raman measurements over ex-situ analysis is unquestionable, ex-situ Raman is useful when used as a preliminary tool to streamline the in-situ measurements, allowing the use of modern methods such as magic angle spinning.[75,76] Given the correct cell design, application to a wide range of state-of-the-art battery designs including Li-S batteries, ASSBs, and thin film batteries.

## 5. In-situ ATR-FTIR for Probing Electrolyte Decomposition at the Electrode/Electrolyte Interface

Fourier transform infrared spectroscopy (FTIR) is based on the absorption of infrared light over the wide spectral range of 14,300 – 20 cm$^{-1}$. In most battery systems, the solvents used as electrolytes are dominated by chemical bonds with energies in the IR regime. Light absorbance as a function of wavelength is calculated to produce a series of absorption peaks. The peaks are caused by excitations of molecular vibration, rotation/vibration, lattice vibration modes, or a combination of each. Each corresponding frequency can be correlated to specific functional groups, while infrared peak intensity can be linearly correlated to the amount of species in the sample.[77] One of the key issues which arises in battery performance assessment is the monitoring of organic species, particularly at the SEI layer. Widely used techniques such as X-ray photoelectron spectroscopy (XPS) provides little to no information on organic SEI species and electrolyte solvents. NMR provides more information on organic solvents but cannot be incorporated to probe the solvents at the surface of the electrode and monitor the SEI layer. In-situ/operando FTIR spectroscopy can be employed at the electrode surface and proves very powerful in the analysis of organic species in real-time, identifying the composition and structure of species within the cell based on the identification of the functional groups present. One of its main uses being the monitoring of electrolyte solvents, including the gaseous products formed upon oxidation and reduction of the electrolytes near the electrode interface.[78] Information can then be provided on the electrochemical changes which occur in the solvents during cycling such as solvation and desolvation.[65,79] Using this information, an insight into the impact that these processes have on the electrodes and species in the vicinity of the electrodes such as the SEI layer can be obtained in real-time, some of which are schematically represented in Fig. 11.[80,81] The FTIR setup is always done in reflective mode, meaning analysis is based on the specular reflectance of the sample rather than simply measuring the energy that passes through. As a result, high sensitivity surface measurements can be acquired without cell disassembly.

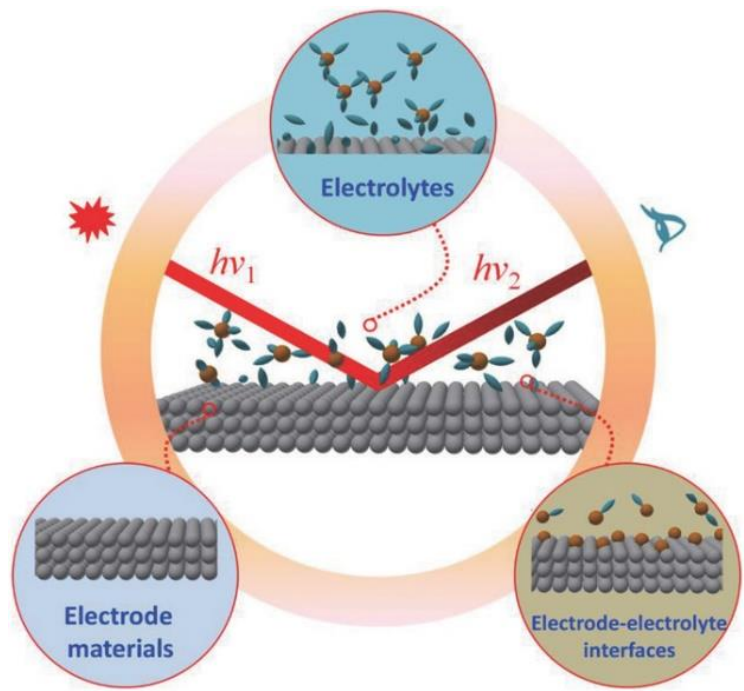

**Figure 11.** Schematic of conventional spectroscopic methods in secondary battery studies focusing on the electrode/electrolyte interface. Reproduced from ref. [82]. Copyright Wiley 2018.

Like the other techniques described thus far the most challenging aspect of using in-situ FTIR to monitor cell degradation is the cell design. With the correct cell design, in-situ FTIR spectroscopy can be applied to a range of battery systems. In the analysis of battery materials, the technique serves mainly to provide insight into the Li$^+$ dynamics at the electrode/electrolyte interface. However, measurements are more applicable to bulk materials, where the penetration depth of the IR beam varies with wavelength. For thin film battery materials, a variation of the technique is more useful, known as attenuated total reflectance Fourier transform infrared (ATR-FTIR) spectroscopy.[83] A schematic of a representative setup is shown in Figure 12 (a). The method operates based on the principle of total internal reflection (TIR). In contrast to traditional FTIR spectroscopy, measurements are independent of sample thickness, and surface sensitivity is enhanced. The penetration depth varies between 0.2 µm and 0.5 µm,[84] with its precise value dependent on wavelength, incident angle, and the refractive indices of the material being probed.

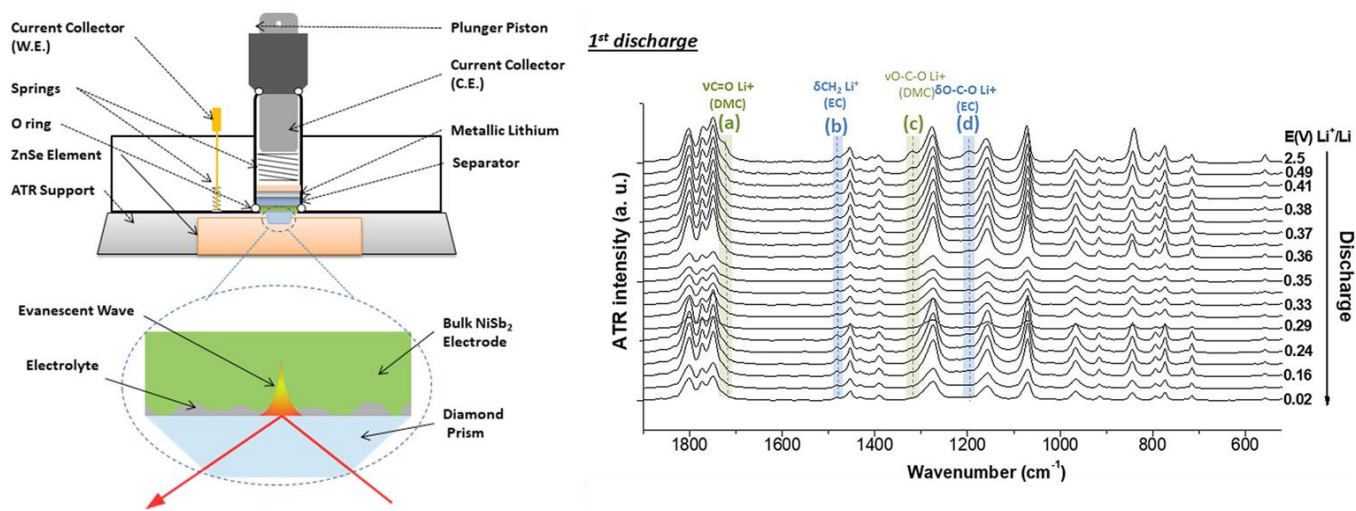

**Figure 12.** (a) Schematic of the electrochemical cell used for the *operando* ATR-FTIR experiment. (b) *Operando* IR spectra obtained during the 1st discharge of NiSb$_2$ electrode at a rate of 1 C with 1 M LiPF6 in EC: DMC (1:1) + 2 % vol VC electrolyte. Adapted from ref. [85]. Copyright ACS 2017.

The high surface sensitivity of the method and its dependence on multiple light reflections mean that it is extremely useful for probing composite and multi-layered materials such as those used in battery systems, remaining useful irrespective of the battery chemistry. In Marino et al., ATR-FTIR spectroscopy is used to probe conversion-type electrode materials.[85] These materials demonstrate interesting performance in terms of capacity but are associated with poor Coulombic efficiency. Poor Coulombic efficiency is usually related to the relentless formation and desolvation of the SEI during conversion and back-conversion reactions. The setup is shown in Figure 12 (a), with a diamond prism used as a waveguide for the ATR-FTIR approach, an NiSb$_2$ electrode and LiPF$_6$ electrolyte dissolved in ethylene carbonate (EC) and dimethyl carbonate (DMC). The measurements are obtained in the vicinity of the electrode/electrolyte interface.

Figure 12 (b) shows the corresponding IR spectra obtained at a rate of 1 C during first discharge. By examining the IR spectra and comparing to known values in the literature, the origin of each of the peaks from species present in both the electrode and the electrolyte. The shift in the peaks of the spectra during cycling means that dynamic evolution of the lithium concentration in the electrolyte can be followed. Tracking of the formation and desolvation of the SEI layer is also achieved. The operando measurements confirm the phase separation of the conversion materials during discharge, and the nanostructuring of the electrode.

Mentioned repeatedly in this review is the promising performance of silicon electrodes and their applicability to next generation battery systems. The ATR-FTIR approach is applied to thin-film amorphous silicon electrodes in operando in Corte et al.,[86] providing information into the characteristic performance issues related to the reaction of silicon to lithiation and the composition of the SEI layer formed. ATR-FTIR assesses the thickness of the SEI layer, which increases during lithiation and partially reduces during delithiation. Question marks remain over applicability of the approach to silicon-based electrodes since they are commonly in (nano)dispersed form, included in a composite with additives and binder, given thar ATR-FTIR spectroscopy is most useful with thin film materials. Nevertheless, the study improves understanding of the processes which govern silicon lithiation and can be applied to any thin film electrode material that does not demonstrate excess IR absorption. This issue often arises in the study of pure metals. As a result, metal oxides are more suitable for IR analysis.

ATR-FTIR spectroscopy proves useful in its ability to probe causes of battery electrode degradation by analysing SEI formation on the electrode interface during cycling. While its use is demonstrated here for the analysis of silicon electrode-based systems,[83] the technique can be applied to several battery technologies even beyond lithium-ion, such as lithium-oxygen, sodium-ion and potassium-ion.[87-89]. A compendium which reviews optical spectroscopy as a tool for battery research can be found in Köhler et al., with a particular focus on UV-visible and IR spectroscopy.[90]

## 6. Electrochemical-Acoustics – Listening to a Battery's Health

In recent years, electrochemical-acoustic methods have been used to monitor the properties of battery materials, particularly SoC and state of health (SoH). Except for NMR, methods of battery assessment discussed thus far have involved the interaction between the sample of interest and some form of electromagnetic energy – X-rays, UV-vis range photons, and infrared light. One of the main disadvantages common between these techniques is their applicability to electrolytes and gaseous species due to their poor scattering properties.[91,92] Acoustic measurements are obtained by applying a series of ultrasonic pulses to the system which can be monitored non-destructively in real-time. Evidence points towards a relationship between ultrasonic transmittance parameters and battery SoC/SoH.[79,93,94] Ultrasound transmission is highly sensitive to gas, porosity, and the mechanical properties of materials. The general approach used to probe battery SoC includes examinations of the relationship between ultrasonic transmittance and the evolution of some of the key physical parameters of the materials.[78,95,96] Changes in the physical dynamics of the system affects the electrochemical activity in the cell and ultimately determines both SoC and SoH.

Operando acoustic time of flight (ToF) modelling and experiments are employed in in Hsieh et al. to correlate structural changes to electrochemical performance.[97] The overarching advantages of the technique are in its simplicity and universal applicability. A combination of modelling and experimental measurements creates a framework relating the distribution of density within a battery system to its SoC and SoH. A representative battery stack used for modelling is shown in Figure 13 (a), with the same setup used for experimental measurements. The setup consists only of the battery and two transducers, one operating in pulse/listen mode and the other only in listen. Unlike the other techniques discussed in this review, the effectiveness of acoustic measurements are independent of the battery chemistry or form factor. The SoC-density relationship can be interrogated since density within the cell affects the speed of the penetrating ultrasonic waves. The speed of sound through a solid material is a function of the elastic modulus and density, given by the Newton-Laplace equation $c = \sqrt{\frac{E}{\rho}}$. Therefore, these two properties govern the speed of the ultrasonic waves which travel through each layer of battery material. A description of the computational model used to simulate the battery stack can be found elsewhere. Strong correlations between SoC and distribution of density within the cell are shown. Furthermore, SoC-density correlations indicate underlying physical processes which occur in electrodes during cycling. A layer-by-layer examination describes mechanical degradation (and evolution) within the cell. Figure 13 (b)-(i) summarise the results of the study, with a clear correlation between the acoustic and electrochemical properties of the $LiCoO_2$/graphite pouch cell during galvanostatic cycling.

Figure 13 (e) shows the sum of all transmitted and reflected waves received by the transducers in green and red, respectively. As the battery is discharged, acoustic absorption increases, which results in a decrease in transmitted and reflected intensity. An exception occurs at end of discharge. The authors attribute this to the capacity-limited cathode. Near 0 % SoC (full discharge), $LiCoO_2$ undergoes a hexagonal-to-monoclinic phase transformation, which dramatically changes both the density and bulk modulus of the cathode, leading to a spike in both the transmitted and reflected acoustic intensities.

Cyclic voltammetry and differential capacity plots can provide an electrochemical fingerprint of a battery system. Light transmission and reflectance spectra can provide an optical fingerprint of battery electrode materials. Here, it is shown that ultrasonic measurements can provide an acoustic fingerprint of the battery system under analysis in real time using a simple methodology in a non-destructive fashion.

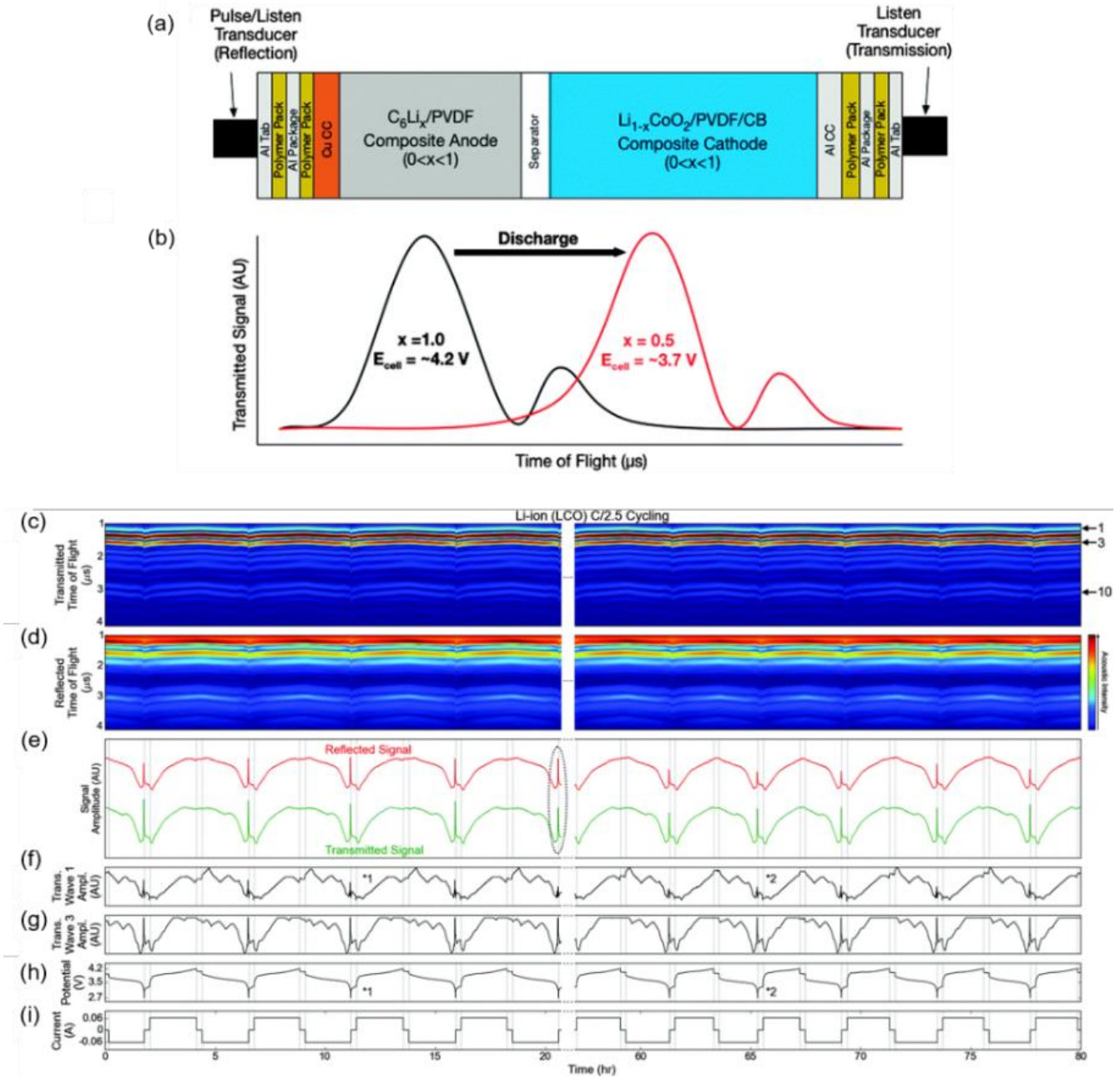

**Figure 13.** Ultrasonic probe of a representative battery. (a) Schematic representation of the battery stack used for acoustic modelling and experimental measurements, shown with packaging, current collector, electrode, separator layers and two acoustic transducers (b) Example illustration of the increase in time of flight (ToF) of the transmitted signal as a function of SOC that occurs during discharge; this shift is a result of the changes in electrode densities as the SOC (i.e., Li content, x) changes. Acoustic behaviour of a LiCoO$_2$/graphite prismatic cell. (c),(d) ToF maps for transmission and reflection modes, respectively, (e) total reflected (red) and transmitted (green) signal amplitudes, (f),(g) traces for the amplitudes of transmitted waves 1 and 3, respectively, (h) cell potential, and (i) applied current as a function of cycling time. The vertical grey lines in panels (e)-(i) represent transitions between charge, discharge, and rest steps. Adapted from ref. [98]. Copyright IOP Publishing 2020.

The acoustic fingerprint is denoted by the ToF of the waves which occur in discrete materials based on the density within the cell. Change in lithiation and therefore SoC alter the acoustic fingerprint, changes which with further research can potentially be quantified based on the change in density and modulus of the material. Currently, there are significant limitations to the model. Many assumptions are made, such as the exclusion of many known non-linear processes which occur during cycling, along with more cell layers.[99,100]

Beyond SoC, density changes indicate underlying physical processes affecting electrode degradation. ToF echo profiles and acoustic signal amplitudes shown as a function of cycle number are key indicators of battery SoC and SoH, indicating inter particle and intra particle stress and strain along with alterations to layer composition of the cell. Physical correlations for large-scale, complex, commercial batteries which have only been probed using high energy X-ray techniques to date are achieved. The method is a fraction of the cost of photonic analysis and negates the need for electrical contact.

Ultrasonic transmittance is particularly useful for detecting the wetting and "unwetting" of Li-ion cells as the presence or absence of the electrolyte at any given position has a significant effect on the ultrasonic transmission at that point. "Unwetting" refers to the swelling of an electrode stack to the point where there is insufficient electrolyte to completely fill the expanded pore space. A phenomenon which occurs during lithium intercalation of IO electrodes during LIB operation. Unlike the techniques discussed previously where initially the techniques are applied ex-situ and then the required adjustments (e.g., cell design) are made to obtain in-situ measurements, the ultrasonic transmission method is in-situ (and operando) by nature. In this case, the advancements that have been made to the analytical power of the technique have been in its spatial resolution. Earlier works employed a fixed-point method, where the ultrasonic transmittance was measured using a beam focused on a particular point or multiple successive points across the cell, providing results averaged over the area of the cell.[101,102] These approaches resulted in a lack of spatial resolution as the beam width was too wide to probe the structural features of the microscale structures which are formed and changed during cycling of a Li-ion cell. One route that has shown success is the use of ultrasonic measurements to identify defects within battery systems and use complementary techniques to analyse the properties of the defects. Robinson et al. used ultrasonic time-of-flight analysis to identify defects and SoH in a Li-ion pouch cell, by examining the acoustic transmission signal through the cell and X-Ray tomography was used to determine the location and scale of the defects.[103] Several studies have taken a similar approach, using ultrasonic transmittance measurements to probe a certain physical/structural change in the material which was then analysed using complementary techniques.[102]

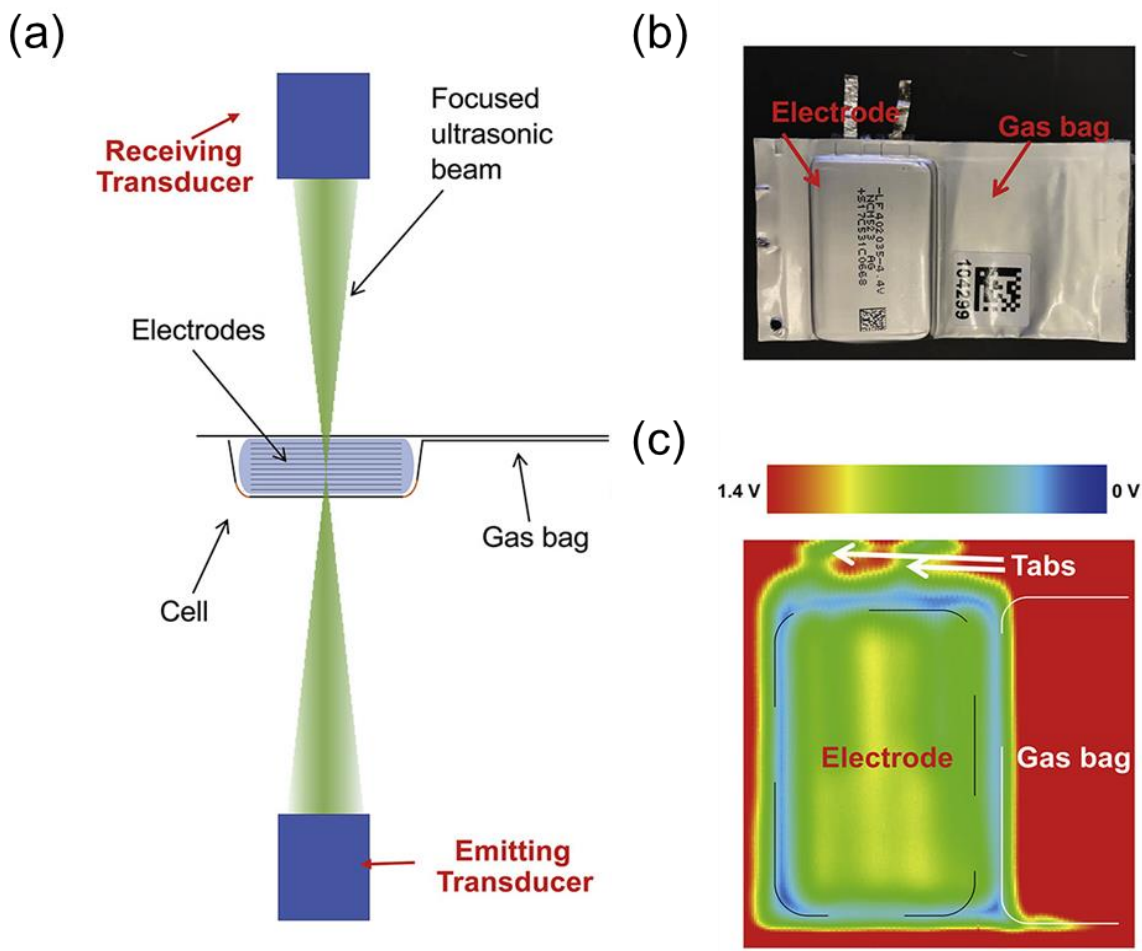

**Figure 14.** (a) Schematic of ultrasonic transmission setup showing path of focused beam. (b) Photograph of pouch cell and its main components. (c) Ultrasonic transmission image of the pouch cell overlapped on the optical photo. Adapted from ref. [39]. Copyright Joule 2020.

A state-of-the-art approach to operando ultrasonic analysis of the Li-ion cells was taken in Deng et al., with an ultrasonic scanning approach that employed a narrow beam with a diameter less than 1mm that provided sub millimetre resolution. The beam was scanned across the cell with a position control accuracy of 0.2mm.[104] Here, complementary techniques were not required. Figure 14 (a) and (b) show a schematic and photograph of the setup used for the ultrasonic scanning of electrolyte wetting and "unwetting" in a Li-ion pouch cell.[62] Gassing, and "unwetting" can be observed in aged cells, which helps to explain their loss in capacity after extended testing. A full map of the ultrasonic transmittance throughout the cell is shown in Figure 14 (c). The ultrasonic scanning technique developed here is cheap, fast, non-destructive and can be applied to large pouch and prismatic cells to monitor the degradation of Li-ion cells on an industrial scale.

Ultrasonic transmittance measurements can be used to monitor the electrochemical-acoustic behaviour and the underlying physical processes can be interrogated in real-time using a cheap, simple, non-destructive method that can be integrated easily into electric vehicles and mobile devices. The method constitutes a valuable addition to the arsenal of the battery analyst, providing the ability to listen to a battery's health during operation.

7. **Tracking the Optical Signature of Photonic Crystal Electrodes in Real Time**

The performance and longevity of battery systems are determined in part by the structural and electrochemical changes which occur within the component materials. In comparison to planar deposits of electroactive material, colour-coded, thin-film battery materials enable real-time diagnostics by observing their "structural" color,[105] caused by a phenomenon known as the photonic band gap (PBG).[106-109] When light is incident on a periodic material with a period on the order of the wavelength of incident light, photons within the PBG are reflected, producing the observed structural colour.[110,111] The observed coloration can be correlated to a specific optical signature for each material, so optical measurements can be used to characterise their structure.[112-114] Such measurements have been widely used for battery materials in recent years[115-119] and are extremely useful for porous electrode structures such as photonic crystals (PhCs).[80,91,120-124] Because periodically porous materials can be fashioned from self-assembly methods through to electron beam lithography and many methods in between that give exquisite control over complexity, the wider research community can grow materials with a tuneable PBG without the need for expensive equipment or complex fabrication techniques.[125] An adjustable PBG in a material gives the opportunity to monitor small changes in material properties using angle-resolved reflectance or transmission measurements[126]. We see PhCs in a wide variety of disciplines, including optical[127-129], energy storage[130,131], biological[132-134] and medical devices[135,136]. Such structures can also behave as optical waveguides[137], a refractive index sensors[138,139], a biosensors[135,140], and also provide opportunities for tuning solar absorption[141,142]. More recently, they have been assessed as binder and conductive additive-free structured electrode materials in Li-ion batteries[115-119].

The move away from the traditional and well-understood battery chemistries to more complex redox processes such as alloying and conversion, and the need to optimize more established chemistries, has motivated the development of new analytical operando techniques that allow study of the fundamental mechanisms by which these materials operate, together with the kinetics of these processes.[21,22,26,143-147] Inverse opal (IO) PhCs have been widely researched as anode materials for lithium-ion batteries (LIBs). Composition changes, Li doping, potential-dependent electrochromism, cation insertion rate-dependent dielectric constant modification and the photonic signatures of lithium insertion into an ordered 3D structure can be achieved by making battery electrode materials in the form of IOs.[78] Some studies have analysed the optical reflectance and transmission of IOs.[16,65,149] As a result, there is a large body of data available when choosing potential anode materials. An extensive review of IOs and their applications to energy storage mechanisms can be found elsewhere.[150] Despite this, optical characterization of IO battery electrodes in situ or in operando has not been achieved.

Fundamentally, in spite of some clear parallels between materials science, electrochemistry and photonic spectroscopies, these disciplines have not yet been combined and brought to bear on new electrode designs in emerging alternatives energy storage materials neither for performance-related device improvements[151], nor fundamental assessment of new modalities, analytics or diagnostics of electrochemical materials response – the research community needs to explore interdisciplinary approaches to drive the next innovation in portable energy storage, especially when new materials and new ways of manufacturing batteries, such as 3D printing, become more advanced.[152-154]

Most transmission/reflectance studies of IO anodes have been carried out prior to cell assembly.[65] This is due mainly to the challenge presented by the closed nature of standard electrochemical cells. Like the use of Raman and FTIR spectroscopy, optical measurements require a less intuitive cell design which can allow light to penetrate the cell. Such adaptations have been made in Lonergan et al., where the optical transmission of IO anodes were examined when placed in a range of solvents.[155] The solvents chosen were potential electrolytes, to predict the effects of the electrolytes on the interaction of light with the materials. A transparent "flooded" cell was used in the experiments, a design previously used for electrochemical performance assessment of PhC anodes in excess electrolyte. The transparent cell design provides light penetration while allowing measurements to be obtained non-destructively. While this represents a step in the right direction by studying the materials in the assembled cell, optical measurements in real time are required to provide a better insight into battery performance.

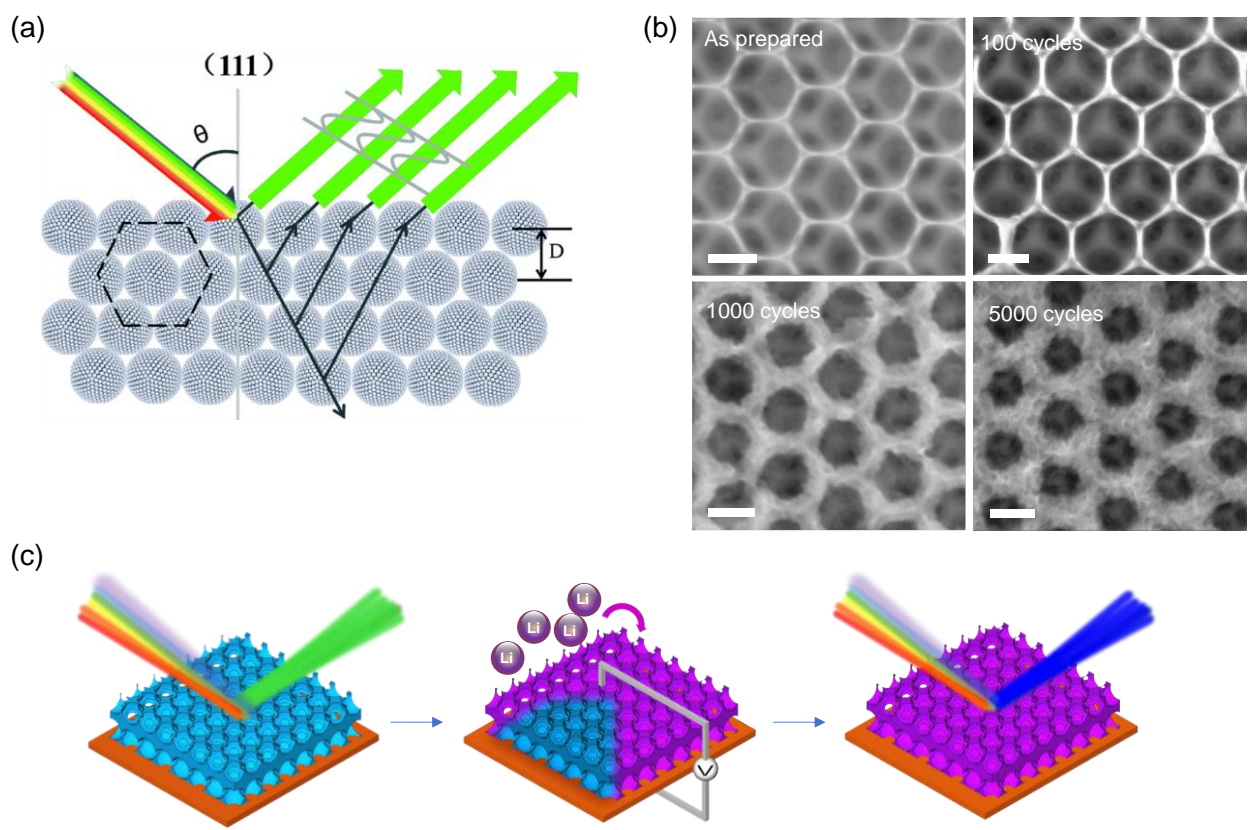

**Figure 15.** (a) Schematic of the PBG in a photonic crystal, selectively reflecting a narrow band of photon energy. Light within the PBG is reflected from the crystal, determined by the lattice constant *D*. Reproduced from ref. [156]. Copyright RSC 2019. (b) $TiO_2$ inverse opal materials used as an anode in a Li-ion battery showing interconnected volumetric swelling as lithiated titanate over 100, 1000 and 5000 cycles. Scale bar = 200 nm. Adapted from Ref. [79]. Copyright Wiley 2017. (c) Schematic representation of operando angle-resolved reflectance measurement of a photonic band gap during reversible lithiation.

3D structured EES materials do not require pristine, defect-free large scale arrays[157] of electroactive deposits as would be required in photonics[158] that use semiconducting inverse opals for optics, telecommunications, optical interconnects, and silicon photonics, as pertinent examples. Optical techniques applied to functioning battery materials in-situ are rare and typically are used to examine the structure of electrolyte compositions of the synthetic preparation of materials[159-161] Only recently, photonic crystals have been used to enhance the photoabsorption characteristics of perovskite halide solar cells using structural colour to tune the absorption across the entire visible spectrum.[162]

While such measurements have not been obtained for IO anodes, operando optical analysis has been achieved for other anode materials.[163] In 2016, Ghannoum et al. used an optical fibre sensor (OFS) to obtain information about the interaction of near infrared light with graphite during its electrochemical lithiation process in a lithium-ion pouch cell.[36] Here, the optical transmission measurements were confined to the overall intensity of the optical signal as a voltage. Further work by Ghannoum et al. in 2020 provided a more extensive optical characterization of the system.[34] As shown in Figure 16, reflectance spectroscopy measurements of lithiated commercial graphite anodes at various states of charge were obtained. To obtain the measurements, the graphite anodes were extracted, and the cells disassembled in an inert environment.[155] Next, an OFS was embedded similar to a previous study,[97] this time in a custom-designed Swagelok cell. Fiber evanescent wave spectroscopy (FEWS)[39] was used to provide operando optical transmission measurements during lithiation of the graphite anode. Reflectance and transmission measurements both demonstrate that there is a direct correlation between the wavelengths reflected/transmitted and SoC.

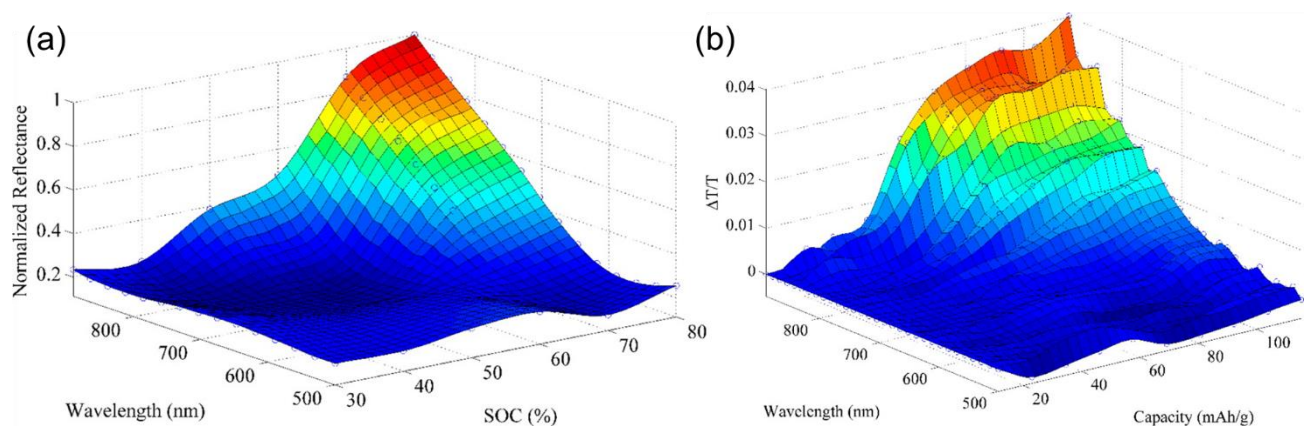

**Figure 16.** (a) In-situ optical reflectance measurements of lithiated commercial graphite electrodes at various states of charge. (b) Operando optical transmission measurements of graphite electrodes during lithiation. Adapted from ref. [149]. Copyright ACS 2016.

The results further highlight the need to monitor these changes in optical activity under operando conditions. After some adaptations, the approach used here may be the key to obtaining operando optical transmission/reflectance measurements of IO anodes.

Tracking the optical signature of PhC electrodes operando may be an effective, low-cost route to probing their degradation mechanisms non-destructively in real-time. Progress is being made in developing the correct methodology for achieving these measurements, with particular attention to consistent cell assembly and internal pressure that does not alter the periodicity of the periodically porous material, nor inhibit a volumetric or other changes that can occur during reversibly ion intercalation. To achieve success, factors such as cell design, light-electrolyte interaction, and electrode response to lithiation must be considered. Like the XRD and acoustic fingerprints observed for battery materials, the optical fingerprint of PhC nanostructures may prove an important asset to further understanding of the physical and chemical processes that govern battery performance. A first step in describing the optical behaviour of ordered porous battery electrodes is a model system to examine charge storage mechanisms, in built stresses, volumetric changes and the intrinsic benefit or scaled porosity to ionic diffusion, electrical conductivity and material interconnectedness, is accurate data in battery electrolytes. For an inverse opal photonic crystal of any materials, such as the example shown in Fig. 15, the following summarise the basic relationships that are useful in characterising the photonic bandgap and its dependence on surrounding environment (air or solvent), effective refractive index and the angle of incidence for reflectance of transmission signal in a typical angle-resolved reflectance/transmission setup.

For a crystal plane (hkl) with interplanar spacing $d_{hkl}$, an IO material with an effective refractive index $n_{\text{eff}}$, some solvent of refractive index $n_{\text{sol}}$ filling the pores, a Bragg resonance order $m$ and some angle $\theta$ between the incident light and the normal to the crystal plane surface, the transmission minimum $\lambda_{hkl}$ can be found via the Bragg-Snell model as follows: $\lambda_{\text{hkl}} = \frac{2d_{\text{hkl}}}{m} \sqrt{n_{\text{eff}}^2 - n_{\text{sol}}^2 \sin^2\theta}$, a plot of the observed minimum wavelength squared versus the sine of the angle of incidence squared should yield a linearly regressive plot with a slope $\mu_0 = -4d_{111}^2 n_{\text{sol}}^2$ and an intercept $\gamma_0 = 4d_{111}^2 n_{\text{eff}}^2$. From this, the interplanar spacing $d_{111}$ and hence the center-to-center pore distance $D$ can be found via the slope. The effective refractive index $n_{\text{eff}}$ of the IO material can be estimated by dividing the intercept by the slope. Currently, there are several methods used in a wide variety of literature to approximate the effective refractive index of composite materials[164,165]. The Drude model for inverse opal systems can be formalized as: $n_{\text{eff}} = \sqrt{n_{\text{IO}}^2 \varphi_{\text{IO}} + n_{\text{sol}}^2 \varphi_{\text{sol}}}$. The Parallel model for an inverse opal system can be written as $n_{\text{eff}} = n_{\text{IO}}\varphi_{\text{IO}} + n_{\text{sol}}\varphi_{\text{sol}}$.

For an inverse opal system, $n_{\text{IO}}$ refers to the refractive index of the crystalline IO material, $n_{\text{sol}}$ is the refractive index of the surrounding medium, $\varphi_{\text{IO}}$ constitutes the volume filling fraction of IO material in the system and $\varphi_{\text{sol}}$ is the volume filling fraction of the surrounding medium. For inverse opal photonic crystal filled with a solvent, a Bragg-Snell relation in terms of the Drude model for $n_{\text{eff}}$: $\lambda_{\text{min}}^2 = 4d_{111}^2(n_{\text{IO}}^2 \varphi_{\text{IO}} + n_{\text{sol}}^2(1 - \varphi_{\text{IO}}))$. Specific details are found elsewhere[95] for a variety of solvent and substrates, but the principle allow, in general, the PBG to be a trackable parameter caused by material and effective index modification when it undergoes electrochemical modification or changes during cycling in a battery cell. Adaption to closed coin cells or similar form factors has yet to be established.

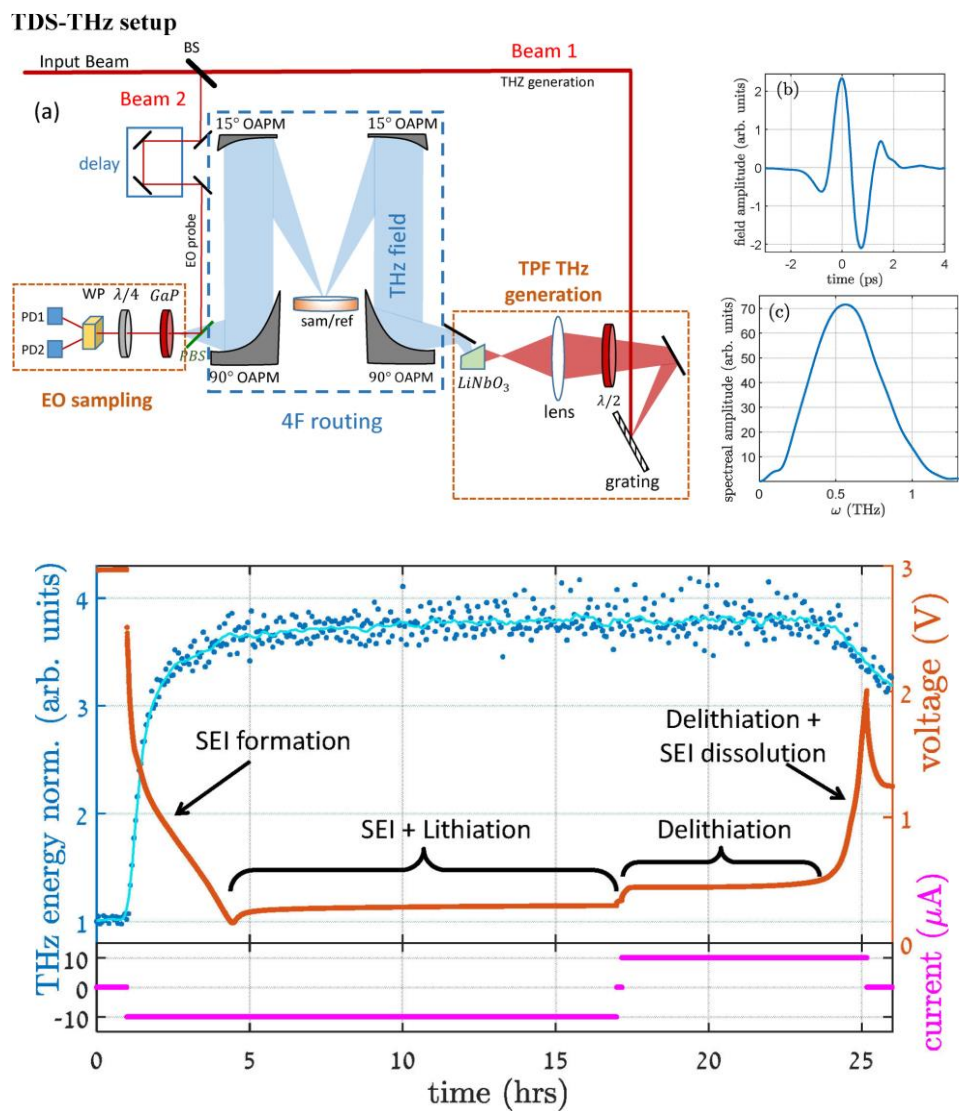

**Fig. 17.** (Top) (a) Schematic of the time-domain THz spectrometer setup in reflection mode with the battery cell placed at prime focus. (b) An example of a THz signal in the time domain together with its associated amplitude spectrum. Electrooptic sampling module includes a pellicle beam splitter (PBS), a Gallium Phosphate crystal (GaP), quarter-wave plate, a Wollaston prism (WP) and two photodiodes. (Bottom) The THz signal (blue dots) in the time domain during a single full cycle. The cycle comprised the discharge (1–17 hours) followed by charging (18–26 hours). The cell voltage is shown in orange and the current in magenta.

Tracking variation in photonic crystal materials in real time spectroscopically allow, in principle, tracking of changes to effective refractive index, periodicity, and conductivity as a function of cycling parameters in voltammetric or galvanostatic conditions. Volumetric swelling and lithiation/sodiation would vary the refractive index contrast and periodicity, shifting the photonic bandgap predictably for a material that is structurally well-defined. In addition, the change or erasure of periodicity, for materials where large cyclic volumetric changes occur, would indicate whether ordered arrangement or in-situ pulverisation and relief of new surfaces is a contributory cause to coulombic efficiency, capacity loss or a limitation in charge rate behaviour for a fixed thickness. These approaches are currently being developed for relevant battery materials to tackle open questions in materials that are assembled with coating morphology and mass loadings that are more relevant to battery cells in various form factors.

Other recent approaches have used operando terahertz time-domain reflectometry spectroscopy[166] (THz-TDS) to investigate solid electrolyte interphase evolution on silicon anodes. THz-TDS operates in the $10^{11}$–$10^{13}$ Hz frequency range[167] as a non-destructive mode. It is based on characterization of the dielectric nature of the sample under test. For battery materials research, THz compatible windows are readily available, and can be integrated into various form factor cells such as coin cells or larger split cells for larger window aperture. The amplitude and phase of the electric field provide the information based on attenuation of the reflected THz field following sample interactions. In this study, Krotkov et al. examined an anode comprising silicon nanoparticles, some binder and a small addition of conductive carbon. In Fig. 17, the basic setup for this approach is summarised along with date from an experiment that allow operando tracking of SEI formation at the surface of the silicon nanoparticles during cycling, with a pronounced dependence on the time between charge-discharge of cycles. They found that certain parts of the SEI tend to dissolve during rest periods if the film was not fully formed prior to the restart of the polarization, particularly for elements of the SEI that are unstable at higher voltages as the charging voltage increases. The measurements show that SEI formation and reformation after dissolution is not only dependent on fresh surface formation that occurs after pulverisation or cracking from repeated volumetric changes. The signal is clear for SEI dissolution following some cracking of the underlying silicon, but that SEI stability at different voltages is not consistent and shown through changes in the effective dielectric constant of the anode during cycling.

## 8. Internal sensing of battery response – Fiber Bragg grating sensors and probes

Motivated by measurement systems in composite research that used Fiber Bragg grating sensors within composite matrices and laminates of materials that allow operando measurement of stress-strain and other mechanical properties[168]. These were developed to avoid some limitation of attachment-type sensors, often based on cantilevers that give uniaxial information (depending on its orientation on the composite or device), and in many cases is limited to device-level information. Recent reports[169,170] have begun to implement FBGs inside battery cells, to get at real-time assessment of the interplay between chemical, mechanical, thermal and other properties and their relationship under specific battery cycling conditions. Fiber Bragg gratins used in this way are assessed by monitoring changes to reflectance or a specific splitting of reflectance peaks caused by birefringence.[171]

Real-time monitoring of the dynamic chemical and thermal behaviour of a battery becomes important when safety and reliability are paramount in new cell chemistries or form factors. Blanquer et al. integrated FBG into 18650 and pouch cells to directly monitor temperature, strain and hydraulic pressure and showed they could be correlated with battery state-of-health and indeed state-of-charge. Careful adjustment of the morphology of the fibre was necessary to enable spectral changes from temperature and pressure in a manner that allowed them to be decoupled.

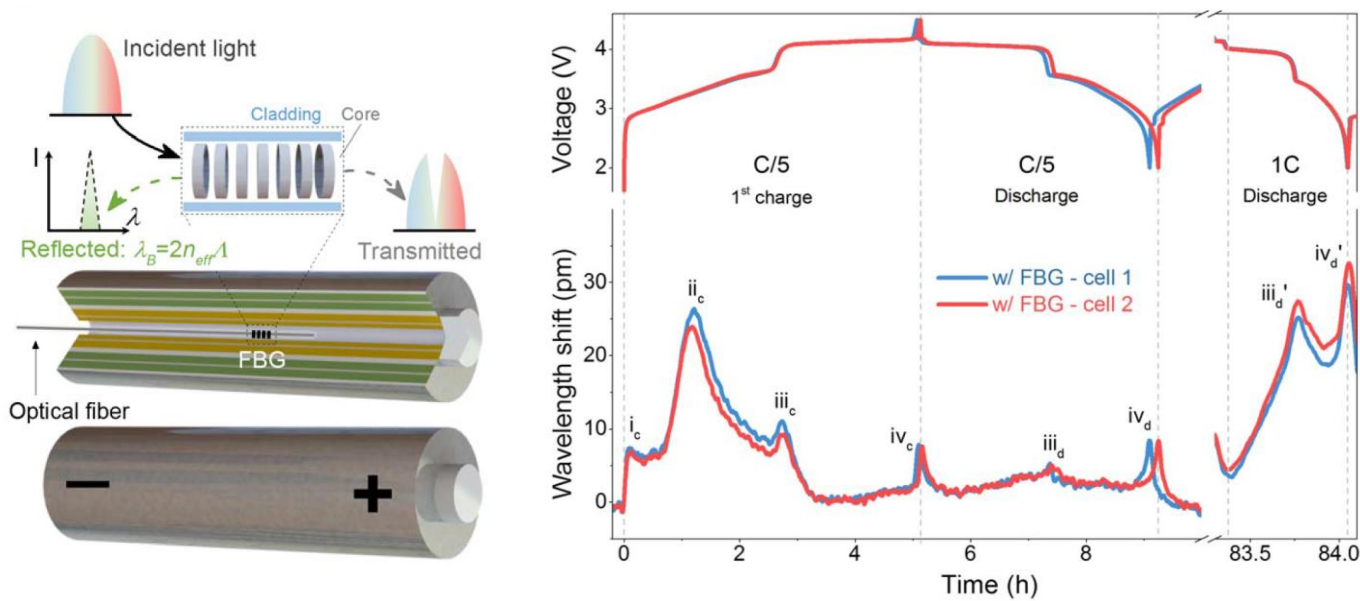

**Fig 18.** (Left) Schematic representation of the integration of optical fiber integration into a 16850 battery cell. (Right) The time-resolved voltage of the cell and spectral shift from the FBG for two separate cells over a single C/5 charge-discharge cycle and also after a discharge at 1 C. Spectral reflectance maxima 1st charge at C/5. The wavelength shift is relative to the wavelength at time = 0 h on the 1st charge at C/5.

This method, summarised in Fig. 18 was then shown to provide operando information on SEI formation and structural evolution of the electrodes. Further details on specific analyses can be found in Ref. [169]. Other work have addressed the possibility of direct SEI formation tracking under operando conditions, notably the work of Louli et al. where operando pressure measurements quantified lithium inventory loss to irreversible cell volume expansion while physically detecting SEI growth during cycling.[172]

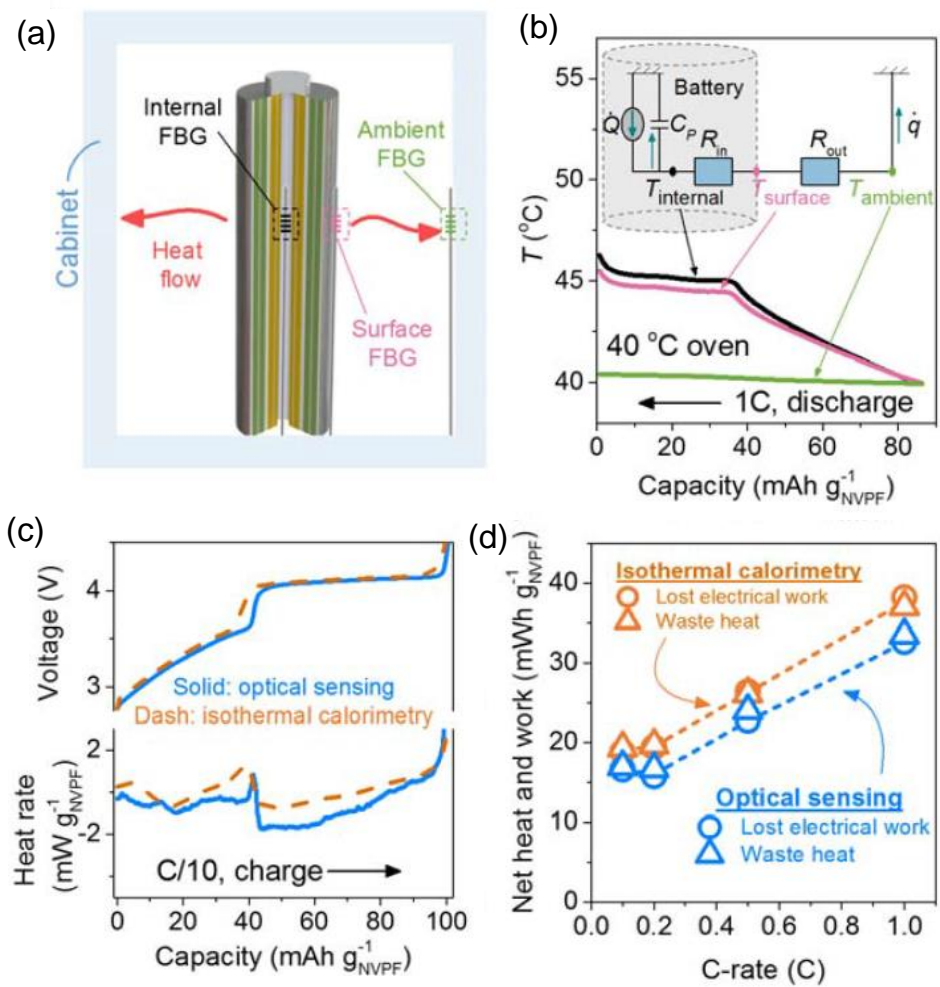

**Fig. 19.** (a) Schematic of an optical sensing calorimetry experiment. (b) The thermal equivalent circuit (inset) and the measured variables including the internal ($T_{internal}$), the surface ($T_{surface}$), and the ambient ($T_{ambient}$) temperatures as a function of time at a discharge rate of 1 C. (c) The comparison of rate of heat generation from the optical fiber method (18650, blue solid lines) and the isothermal calorimetry (coin cell, orange dash lines) measured by the authors at a charge of C/10. (d) Comparison of waste heat and lost electrical work between the FBG approach (18650, blue) and the isothermal calorimetry (coin cell, orange). Adapted from Ref. [170] where more specific details on the cell chemistry and measurement conditions can be found. Copyright Nature Publishing Group, 2020.

The same group also adapted the approach by using multiple sensors (Fig. 19) to measure cell-generated heat curing cycling and compared the operando measurements to those by isothermal calorimetric measurements.[170] One major motivation from a practical standpoint is that current heat generation and dissipation measurement techniques are slow, require larger infrastructure and are limited to measurements outside the cell. The authors detail the basis of their heat mapping models, factoring a zero-dimensional model that was scalable (Fig. 19). Three separate FBGs were used to measure ambient, internal and surface temperatures of the cell by monitoring temperature dependent spectral shifts. Using a benchmark method to quantify temperature variations as a function of time following a galvanostatic pulse to the cell and extrapolation of model parameters. The approach allows the derivation of the rate of heat generated by the cell and heat flow rates under any electrochemical conditions.

Because of the sensitivity and relative accuracy of these approaches, it is conceivable that adaptions may allow for better thermal management systems, modification of electrolytes and other factors bespoke for new cell chemistries during directly measuring such quantities and properties within an actual cell under cycling conditions. And one very practical advantage is the ability to monitor the cell under more aggressive cycling conditions (higher charging rates for example) and the decouple heat transfer and heat accumulation in real-time within 18650 cells, and possibly other form factors too. Another important consideration in some battery materials such as alloying phases with large volume changes during reversible Li reactions, and interfaces in all solid state batteries (ASSBs) is the measurement and quantification of stresses and their relationship to materials and their interfaces. These effects are critical for long life performance, and this is especially important for ASSBs that used solid materials and electrolytes with pristine interfaces and no porous component to buffer and changes in internal stresses. Using FBGs and moving the sensing capability inside the battery has obvious potential benefits for localised measurements compared to sensors located outside the entire cell.

In work by Blanquer et al., FBG sensors installed into Swagelock battery cells were used to directly measure internal stress evolution in liquid and solid-state electrolyte systems. Following Bae et al. whose approach showed operando strain measurements using an optical fiber grating in lithium-ion battery electrodes[173], Figure 20 summarises the overall approach using a liquid-based electrolyte in this case, the spectral shifts measured at specific states of charge and discharge as time-resolved data. The optical signal was monitored during battery cycling, further translated into stress and correlated with the voltage profile. While the details specific to the alloying compound (In, Si) can be found in Ref. [169], we summarise here for readers some of the basic parameters that allow for spectral shift, conversion to stress data and the basis for birefringence observations that underly this technique. Measurements that build on these fundamentals may be possible for certain cell where open questions remain about certain materials, interfaces, or diagnostics.

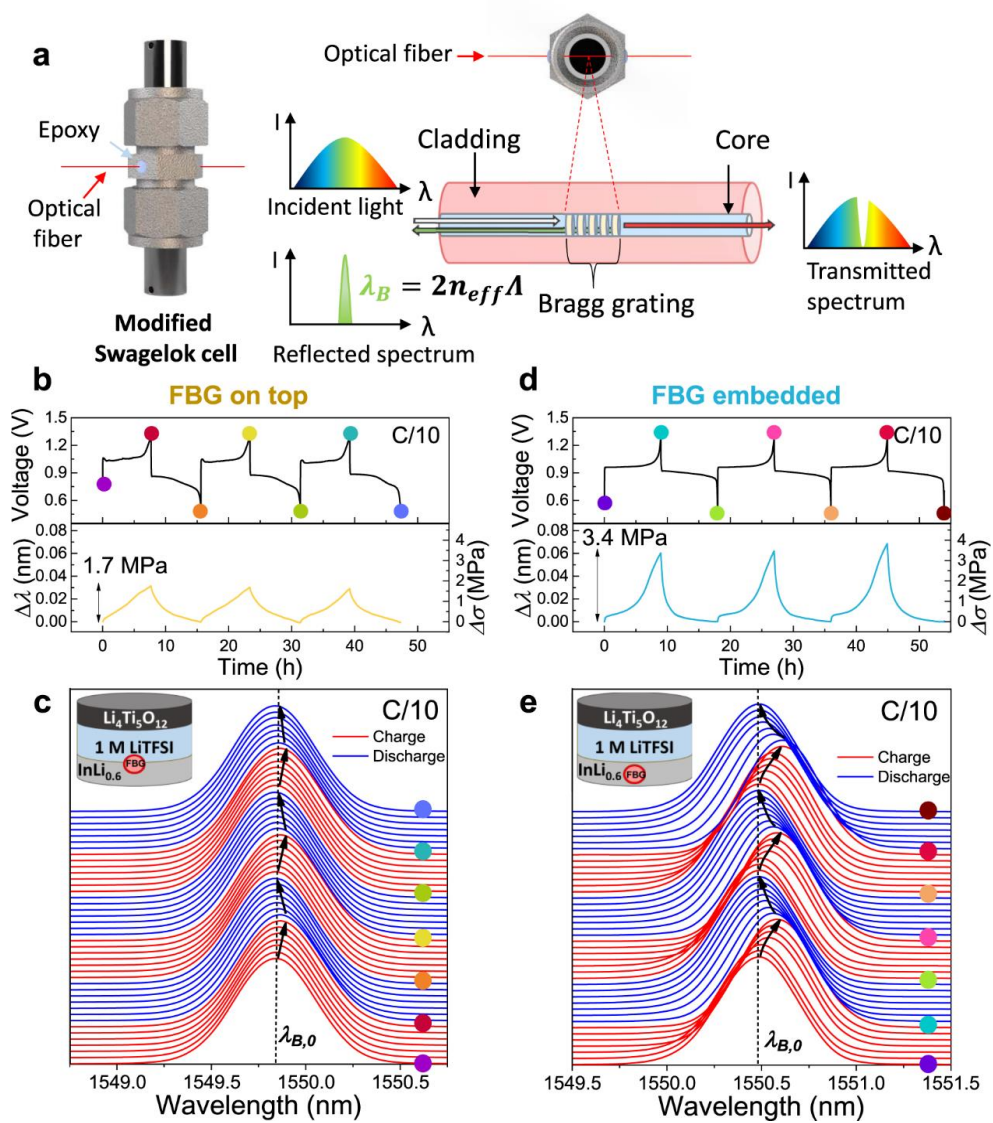

**Fig. 20.** (a) Integration of an FBG into an in-house-modified Swagelok cell with a graphical summary of the working principle of an FBG optical sensor. (b) Time-resolved voltage (top) and $\Delta\lambda_B$ and $\Delta\sigma$ (bottom) evolution from the FBG sensor of an $InLi_{0.6}$ - LTO cell in a 1 M LiTFSI in DOL:DME electrolyte. The FBG was put at the anode/electrolyte interface. (c) Reflectance spectra from the FBG sensor located at the anode/electrolyte interface for the cycles shown in (b). (d, e) Analogous plots to (b, c), for a cell with the FBG sensor embedded within the $InLi_x$ electrode. Adapted from Ref. [169]. Copyright Nature Publishing Group 2022.

This may need new thinking for FBG placement, or indeed other techniques that allow operando tracking of optical responses unique to the materials, such as photonic crystals or periodically patterned electrodes. When light travels through the optical fiber, the FBG sensor acts as a reflector for a specific wavelength, namely the Bragg wavelength (λB) which is defined as $\lambda_B = 2n_{eff}\Lambda$, where $n_{eff}$ is the effective refractive index of the FBG its is immediate environment and $\Lambda$ is the Bragg grating period. Any temperature (T), hydraulic pressure (P), or strain (ε) change detected by the FBG sensor changes the value of $n_{eff}$, $\Lambda$ or both which causes a shift of the reflection maxima. Then, under a condition when the FBG is strained one can describe the shift of $\Delta\lambda_B$ [174-176] as

$$\Delta\lambda_B = \left(1 - \frac{n_{eff}^2[p_{12} - \nu(p_{11} + p_{12})]}{2}\right)\varepsilon$$

where $\Delta\lambda_B$ is the spectra shift compared to the initial measurement, $p_{11}$ and $p_{12}$ are the strain-optical coefficients of the fiber being used together with the Poisson ratio, $\nu$. As such, the longitudinal strain measured spectrally can be converted simply to stress using Hooke's law according to $\sigma = \varepsilon E$, using $E$ which is the Young's modules of the fiber. In some cases, cycling a battery material may involve other directional strains, mimicking a transverse (or non-longitudinal strain) that causes a birefringent reflectance peak splitting[177] from different light polarization and effective index, according to Gafsi et al. [178]

$$B = \frac{|n_\parallel - n_\perp|}{n_{eff}} = \frac{|\Delta n_y - \Delta n_x|}{n_{eff}} + B_0$$

where $n_\parallel$, $n_\perp$, $n_y$, and $n_x$ represent indices for directions related to external force and the index changes due to different polarization. $n_{eff}$ is the base effective index prior to modification. Ther are necessary calibrations required to effective deconvolute peak splitting that are detailed in several publications, and in Ref. [169] for calibration curves obtained for the cell at OCV. In summary, this method provide additional resolution in the analysis of in-situ stresses in longitudinal and transverse directions with represent to the FBG on ASSB electrodes during cycling.

Operando stress measurements using FBGs has significant potential to track chemico-mechanical processes inside electrodes and at their interfaces, both for probing fundamental questions on material behaviour during cycling and as a diagnostic toolset for new materials under various cycling conditions. This is important for ASSBs that rely on well-defined interfaces and tracking stress build up in general will require adaption of the technique to place FBG probes at optimum regions in any form factor cell that could benefit from accurate operando stress build up measurements.

## 9. Battery Operation in a Dynamic Service Environment – Building a Sensor Network

While analysis of individual cells and their components provides information on battery performance, it is also necessary to examine batteries under dynamic conditions. Often, batteries are used as components of larger battery pack systems. In Li et al.,[179] Li-ion battery monitoring was performed operando using a sensor-based network. The network was used to examine the effects of vibration and impact on battery performance during cycling, to provide a simulation of the conditions the system may experience during transport. The relationships of both temperature variance and deformation to capacity degradation during cycling in various dynamic environments are established. The network identifies critical points during cycling where high heat generation and stress accumulation may lead to thermal runaway and potential explosion. The comparison between single cell and battery pack performance is also explored.

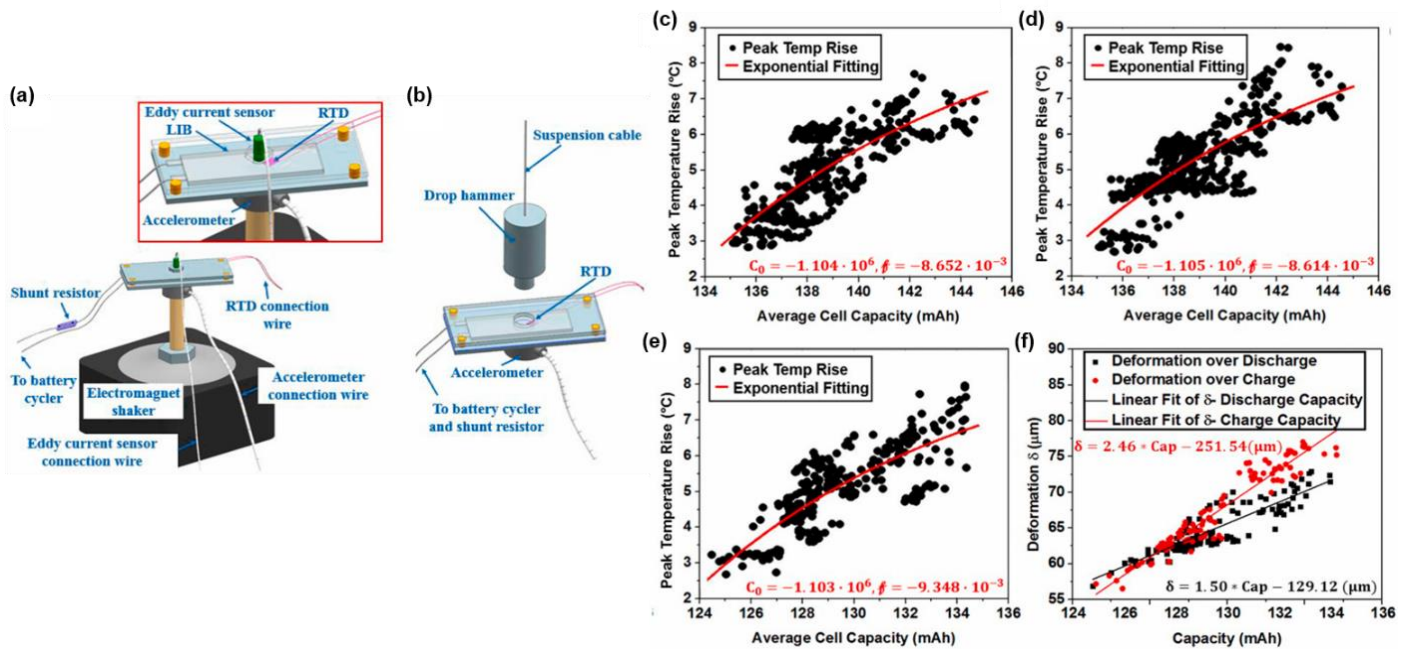

**Figure 21.** Operando testing platform for LIBs: (a) setup for operando vibration test, and (b) setup for operando impact test analysis. Relation between maximum temperature increase and capacity of (c) cell 1, (d) cell 2 and (e) cell 3. Relation between maximum deformation and capacity of cell 3. Adapted from ref. [179]. Copyright Elsevier B.V. 2020.

In Figure 21, the operando testing platform used for (a) the vibration test and (b) the impact test of the LIBs is shown. An electromagnet shaker was used to simulate vibration, while a drop hammer simulated impact. In the sensor network, capacity degradation was measured using the internal resistance temperature detector (RTD) and eddy current sensor. Vibration and impact testing investigated the changes in battery capacity over the typical service life of 500 cycles. The testing results showed capacity retention immediately after impact, with capacity degradation observed after continuous cycling. Low momentum impact can release compressive stress in the electrodes, while high momentum impacts van cause severe cracking, each leading to degradation and capacity loss. The evidence indicates that dynamic loading has little effect on battery performance in the short term, but repeated stress leads to electrode degradation and battery failure.

Cell deterioration and the role of electrode degradation is explored in this work by examining the relationships between capacity and two of the key physical properties which affect it, temperature change and deformation, for both the single cell and 2-cell pack. In Figure 21 (c)-(e), the temperature-capacity relation is shown for each of the three cells. The relation provides an indirect method to measure temperature, eliminating the requirement for dedicated temperature monitoring, reducing the number of components consuming power in the network and reducing cost. However, for a multi-cell battery, the effects of adjacent cells must also be considered. Figure 21(f) shows the deformation-capacity relation. Deformation measurements provide information on structural changes to the cell. Like the work discussed throughout this review, this study links structural changes to the electrochemical processes which occur during cycling, via operando analysis. For the sensor network designed here, deformation indicates the level of stress accumulation during service which can lead to dendrite formation causing internal short circuits. The evidence shows greater deformation occurs during charge than during discharge, which causes the cell to swell over its cycle life. The overall deformation growth can be attributed to several factors. These factors can be broken down into two categories, gas generation and electrode thickness change. By mapping deformation to electrode thickness change, the role of electrode degradation to cell performance can be established. The sensor network can not only detect degradation but identify its source. Here, the results show that deformation is greater than thickness change, which occurs only in the graphite anode and not the LCO cathode. The results demonstrate that anode degradation and gas generation (due primarily to electrolyte decomposition), is the key factor leading to deformation of the cell.

As the inevitable move towards renewable energy gains momentum, energy storage requirements mean that battery performance assessment methods must not be left behind. The use of a sensor network as described above which can pinpoint performance issues in a multi-cell configuration can support the employment of battery systems at the grid level. The methods outlined in this short review highlight how many cross-disciplinary studies are now being conducted to benefit the electric future, from materials choice and abundance, to long term high performance battery response, while in parallel gaining better understanding on issues and safety concerns from new battery chemistry and designs to detailed microstructure, interface and localised reactions that govern, control improve and impede safe, efficient cyclic charging and discharge over a long lifetime.

**Conclusions and outlook**

Many of the achievements using operando XRD, NMR and synchrotron techniques[180] have provided useful capabilities and new insights in terms of operando detail and understanding of critical processes in batteries and electrochemical materials science. However, they do not readily translate into real-time monitoring of batteries in the field because of the nature of the infrastructural requirements. These methods do provide very detailed and powerful ways of assessing material and cell-chemistry behaviour in real time. Methods that will allow detection, in real time, of the complex combination of intended chemical and electrochemical processes, and of the panoply of side reactions that significantly affect long-term performance will be useful and compliment synchrotron-based methods when used for material prediction, screening and safety diagnostics. This will allow the material science, chemistry, electrochemistry, and battery communities to optimize cycling regimes or identify failure mechanisms more rapidly in emerging technologies.

Battery history can then be determined and a health 'check-up' to be performed, allowing appropriate pricing and insurance for a battery as it enters its second-life application. Grid and transport demonstration projects and battery tests are already in progress to link traditional electrochemical responses (such as current, voltage and impedance) with test routines appropriate to the technology and produce the 'big data' needed to extract new correlations and, ultimately, predict future performance. Current examples at the time of writing are large scale European projects under the Battery2030+ initiative that is focused on real-time, passive, and other methods to monitor battery performance, safety, and cell health.  An inside-the-battery-material analogue of today's battery management systems would prove useful too. Deriving useful information from the thousands of batteries used commercially via data-driven consumer apps and coordinated mapping of testing procedures from lab-based tests of new systems are ways being investigated now to obtain a big-data map of all that affect batteries and battery components. The details of these larger scale, forward looking initiatives are outside the scope of this review and the reader is directed to available internet resources on these initiatives[181-187]. Diagnostics are required that can couple electronic with chemical traceability and can identify key reactions that are critical to both short- and long-term cell failure.

These methods will be vital for Na-ion, solid state batteries and related battery formulations, as well as many electrochemical systems where solid-solution interfaces are common and where dynamic processes in materials control and affect the nature of their response and behaviour. Li-ion technologies are now proven from data to in principle power an EV for a 'million miles', a reference to seminal work by Dahn and co-workers[188] where crystalline NMR532 cathodes and balanced artificial graphite cell were demonstrated. The cells used in that study were never turned off and are still cycling at the time of writing to many tens of thousands of cycles within the optimized voltage window. The same group have also demonstrated that an ionic liquid electrolyte and choice voltage window can allow similar cells to run for effectively more than a century at 25°C, and state of the art stability and longevity at extremely high (~70°C and higher) temperatures.[189] As Na-ion make inroads to be the next replacement for applications where Li-ion is critical now, screening, prediction and real-time analysis will be essential in its future. Safety matters will always be obviously important and leveraging many of these techniques as predictive or real-time monitoring methods in test and also in commercial battery systems with communication functionality to the user will be key components that help battery technology users, such as EV drivers, to better understand the battery and its health, as much as has been learned by users of ICE equivalents.

Ultimately, big-data analysis and prediction of battery behaviour and new materials, together with large-scale high resolution operando methods with built-in analytical sensors in consumer batteries will form a trifecta of analysis that can help to optimize existing and future battery use and development.

**Acknowledgements**

We acknowledge support from the Irish Research Council Advanced Laureate Award under grant no. IRCLA/2019/118.